\documentclass[fleqn,usenatbib]{mnras}

\usepackage{newtxtext,newtxmath}

\usepackage[T1]{fontenc}

\DeclareRobustCommand{\VAN}[3]{#2}
\let\VANthebibliography\thebibliography
\def\thebibliography{\DeclareRobustCommand{\VAN}[3]{##3}\VANthebibliography}

\usepackage{appendix}
\usepackage{lscape}
\usepackage{ulem}
\usepackage[detect-none]{siunitx}
\usepackage{tabularx}
\usepackage{graphicx}	
\usepackage{amsmath}	
\usepackage{natbib}
\usepackage{lineno}
\usepackage[usestackEOL]{stackengine}

\newcommand{\emerlin}{\textit{e}{-MERLIN}}
\newcommand{\microJybm}{\,\mathrm{\umu Jy\,beam^{-1}}}

\newcommand{\plus}{$\pm$}

\title[The Quasar Feedback Survey at 6\,GHz with \emerlin{}]{The Quasar Feedback Survey: zooming into the origin of radio emission with \emerlin{}}

\author[A. Njeri et al.]{Ann Njeri,$^{1}$\thanks{E-mail: ann.njeri@newcastle.ac.uk}
Chris M. Harrison,$^{1}$\thanks{E-mail: christopher.harrison@newcastle.ac.uk}
Preeti Kharb,${^2}$
Robert Beswick,$^{3}$
Gabriela Calistro-Rivera,$^{4,5}$
\newauthor
Chiara Circosta,$^{6}$
Vincenzo Mainieri,$^{5}$ Stephen Molyneux,$^{7}$  
James Mullaney$^{8}$
and Silpa Sasikumar${^9}$   
\\
$^{1}$School of Mathematics, Statistics and Physics, Newcastle University, Newcastle upon Tyne, NE1 7RU, UK\\
$^{2}$National Centre for Radio Astrophysics (NCRA) - Tata Institute of Fundamental Research (TIFR), S. P. Pune University Campus, Post Bag 3, Ganeshkhind,\\ 
411007 Pune, India\\
$^{3}$Jodrell Bank Centre for Astrophysics, School of Physics and Astronomy, The University of Manchester, Alan Turing Building, Oxford Road, Manchester,\\  M13 9PL, UK\\
$^{4}$German Aerospace Center (DLR), Institute of Communications and Navigation, Wessling, Germany\\
$^{5}$European Southern Observatory (ESO), Karl-Schwarzschild-Stra{\ss}e 2, 85748 Garching bei M\"{u}nchen, Germany\\
$^{6}$ESA, European Space Astronomy Centre (ESAC), Camino Bajo del
Castillo s/n, 28692 Villanueva de la Ca\~{n}ada, Madrid, Spain\\
$^{7}$School of Physics and Astronomy, University of Southampton, Southampton, SO17 1BJ, UK\\
$^{8}$The School of Mathematical and Physical Sciences, University of Sheffield, Sheffield, S3
7RH, UK\\
$^{9}$Departamento de Astronom\`{i}a, Universidad de Concepci\'{o}n, Concepci\'{o}n, Chile\\
}

\date{Accepted 2024 December 24. Received 2024 December 19; in original form 2024 November 8}

\pubyear{2025}

\begin{document}
\label{firstpage}
\pagerange{\pageref{firstpage}--\pageref{lastpage}}
\maketitle

\begin{abstract}

We present 6\,GHz \emerlin{} observations of 42 $z<0.2$ type 1 and type 2 mostly radio-quiet quasars ($L_{\rm[OIII]}\gtrsim10^{42}$\,erg\,s$^{-1}$; $L_{\rm AGN}\gtrsim10^{45}$\,erg\,s$^{-1}$) from the Quasar Feedback Survey. The nature and origin of radio emission in these types of sources is typically ambiguous based on all-sky, low-resolution surveys. With \emerlin{}, we investigate radio emission on sub-kiloparsec scales ($\sim$10s-100s\,pc). We find 37/42 quasars are detected, with a diversity of radio morphologies, including compact cores, knots and extended jet-like structures, with sizes of 30--540\,pc. Based on morphology and brightness temperature, we classify 76\,per\,cent of the quasars as radio-AGN, compared to the $\sim$57\,per\,cent identified as radio-AGN at the $\sim$1--60\,kpc scales probed in prior studies. Combining results from \emerlin{} and the Very Large Array, 86\,per\,cent reveal a radio-AGN. On average, $\sim$60\,per\,cent of the total radio flux is resolved away in the \emerlin{} maps, and is likely dominated by jet-driven lobes and outflow-driven shocks. We find no significant differences in measured radio properties between type 1 and type 2 quasars, and estimate sub-relativistic jet speeds of $\sim$0.2-- 0.3c and modest jet powers of $P_\mathrm{jet} \approx \times$10$^{43}$\,erg\,s$^{-1}$ for the few targets, where these measurements were possible. These quasars share characteristics with compact radio-selected populations, and the global radio emission likely traces strong interactions between the AGN (jets/outflows) and their host galaxy ISM from 10s\,parsec to 10s\,kiloparsec scales.  
\end{abstract}




\section{Introduction}
For galaxy evolution models and simulations to fully explain the properties of massive galaxies, Active Galactic Nuclei (AGN) are required to inject energy into their environments. This produces a `feedback' process, which re-distributes baryons and regulates galaxy growth (e.g., \citealt{Bower2006,Hirschmann2014,Schaye2015,Henriques2015,Choi2018,Pillepich2018b}; reviews in e.g., \citealt{Cattaneo2009,Fabian2012,HeckmanBest2014,King2015,Harrison2017,Saikia2022,Harrison2024}). However, the role of powerful `radiatively dominated' AGN (i.e. quasars with $L_{\rm AGN}$ $\gtrsim$10$^{45}$\,erg\,s$^{-1}$) remains controversial from both a theoretical and observational perspective. There is an ongoing debate around the most important mechanism of energy injection from these sources (i.e., jets, accretion-disk winds or direct radiation pressure on the host galaxy gas) and to what level they can have an appreciable impact on galaxy evolution \citep[e.g.,][]{Wylezalek2018,Cresci2018,Harrison2024}.

Observations show that understanding the origin of radio emission in typical quasar host galaxies is a key piece of the AGN feedback puzzle. For example, intriguing trends between radio emission, dust redenning, and/or the prevalence of outflows, have been presented throughout the literature for radiatively luminous AGN (e.g., \citealt{Mullaney2013,VillarMartin2014,Zakamska2014,Hwang2018,Glikman2022,Fawcett2023,CalistroRivera2024,Petley2024,Kukreti2024,Liao2024,Alban2024}; although also see \citealt{Wang2018,Rakshit2018}). However, since most quasars have moderate radio luminosities \citep[i.e., $\gtrsim$90\% with $L_{{\rm 1.4GHz}}\lesssim10^{25}$\,W\,Hz$^{-1}$;][]{Zakamska2004} and have radio structures that are typically unresolved in all-sky surveys, it is often unclear what dominates the radio emission in this population. The primary candidates are: compact radio jets (i.e. $\lesssim$ a few kiloparsec); shocks from AGN-driven outflows; star formation processes, and/or accretion disk coronal emission \citep[][]{Kellerman1994,Laor2008,Condon2013,Mullaney2013,Padovani2016,Behar2018,Panessa2019,Harrison2024}. Indeed, at typical sub-mJy flux density levels of radio emission, only a small fraction of quasars are unambiguously identified as AGN, with the bulk of their radio emission associated with either star formation \citep[e.g.,][]{Kellerman2016,Herrera2016} or both star formation and AGN processes \citep[e.g.,][]{Silpa2020,Wang2023}.

The most commonly observed radio morphology in large radio surveys are single compact objects \citep[e.g.,][]{ODea1997,Tedhunter2016,Sabater2019,Chiaraluce2019,Fawcett2023}. If AGN dominated, the compact nature of these objects may imply intense interaction with the interstellar medium ISM at parsec to a few kiloparsec scales, either through small-scale radio jets or shocks caused by outflows (launched by accretion disk winds, radiation pressure or jets; e.g., \citealt{Mullaney2013,Zakamska2014,Mukherjee2018,Bicknell2018,Saikia2022,Liao2024,Petley2024,Chen2024c,Harrison2024}). Recent high resolution radio observations reveal a high prevalence of compact radio jet-like structures at sub-kiloparsec and parsec scales for lower radio luminosity AGN and quasars \citep[e.g.][]{Hardcastle2019,Hardcastle2020,Boccardi2021,Jarvis2021,McCaffrey2022,Njeri2023}. Furthermore, studies of individual AGN-host galaxies, combining high resolution radio imaging with multi-phase gas kinematics, reveal a common spatial connection between compact radio structures, covering a wide range of powers, and multi-phase AGN-driven outflows \citep[e.g.,][]{Rosario2010a,Husemann2019,Morganti2013,Riffel2014,Tadhunter2014,Jarvis2019,Santoro2020,Venturi2021,Venturi2023,Girdhar2022,Girdhar2024,Cresci2023,Speranza2024,Ulivi2024,Holden2024}. This may all indicate a mechanism of AGN feedback acting in moderate radio power systems, as an alternative to that typically associated with the mechanical input caused by powerful relativistic radio jets on the largest spatial scales \citep[e.g.,][]{Bridle_Perley1984,Laing2015}. 


{\bf The Quasar Feedback Survey:} Recent work has clearly shown the importance of studying radio emission to understand AGN feedback \textit{across a wide range of radio powers}, with at least some of the radio emission tracing an ongoing interaction between the AGN and the host galaxy multi-phase interstellar medium. Nonetheless, key questions remain, including: how common is such an interaction for typical quasars? what is the efficiency of energy transfer from the AGN to the ISM?; what is the long-term impact on the host galaxies?; and what fraction of the radio emission can be associated with various processes (shocks, radio jets, and star formation). Towards addressing these outstanding questions, we are undertaking a systematic survey of low redshift $z<0.2$ quasars. We have constructed a sample of 42 $z$\,$<$0.2 quasars ($L_{\rm AGN}>10^{45}$erg\,s$^{-1}$) selected on their [O~{\sc iii}] luminosity, and which is dominated by typical `radio quiet' AGN \citep[][]{Jarvis2021}. The low redshift sample makes it possible to study in high detail the physical interaction between the AGN and their host galaxies. Using a representative sample of quasars, the survey is designed to characterise the multi-phase outflow properties and dominant outflow driving mechanisms \cite[][]{Harrison2014,Lansbury2018,Jarvis2019,Girdhar2022,Girdhar2024}, assess the impact of the AGN on the host galaxy properties \citep[][]{Jarvis2021,Molyneux2024} and to characterise the origin and properties of the radio emission \citep[][]{Jarvis2019,Jarvis2021,Silpa2022}. The latter is the focus of this current study. 

We have already obtained C-band (6\,GHz) and L-band (1.4\,GHz) Karl G. Jansky Very Large Array (VLA) observations of these 42 targets and produced radio maps down to a resolution of $\approx$0.3$^{\prime\prime}$ \cite[i.e., $\sim$1\,kpc; presented in][]{Jarvis2019,Jarvis2021}. Although we identified morphological radio structures, on $\gtrsim$\,kpc scales, many of these structures were ambiguous in origin. For example, they were often not clearly collimated into jet-like structures and/or the majority of the flux was found to be located in unresolved/featureless nuclear regions, for which we were unable to distinguish between an AGN or star formation origin. To assess the properties of the radio emission on sub-kiloparsec scales, for this study, the 42 VLA quasars have been followed up with $\sim 6\,$GHz observations on the enhanced Multi Element Remotely Linked Interforometer Network (\emerlin{}). \emerlin{}\footnote{\url{https://www.e-merlin.ac.uk}} is an array of seven radio telescopes spanning 217\,km across Great Britain, making it possible to reach $\sim$50\,mas resolution at $\sim 6\,$GHz. 

This paper is organised as follows. In Section~\ref{sec:obsvns} we describe the observations, imaging strategies and data reduction. Section~\ref{sec:analysis} presents the \emerlin{} results and analysis. Section~\ref{sec:discussions} presents a discussion on the radio properties and likely origin of the radio emission, across our quasar sample. This includes information from the previous VLA observations and how the sample may be related to compact radio AGN populations in the literature. Our conclusions are presented in Section~\ref{sec:conclusion}.  In this paper, we adopt the $\Lambda$CDM cosmology parameters with $H\mathrm{_0} = 70.0$\,kms$^{-1}$\, Mpc$^{-1}$ and $ \Omega_{\rm m} = 0.30$. In this cosmology, 50\,mas (\emerlin{} resolution at $\sim 6\,$GHz), corresponds to 130\,pc for the median redshift of the sample ($z$=0.15). The spectral index, $\alpha$ uses the convention $S_\nu \propto \nu^{\alpha}$, where $S_\nu$ is the integrated flux density. 


\section{Observations and data reduction}\label{sec:obsvns}

\subsection{Sample selection}
\begin{figure}
\centerline{\includegraphics[width=\columnwidth]{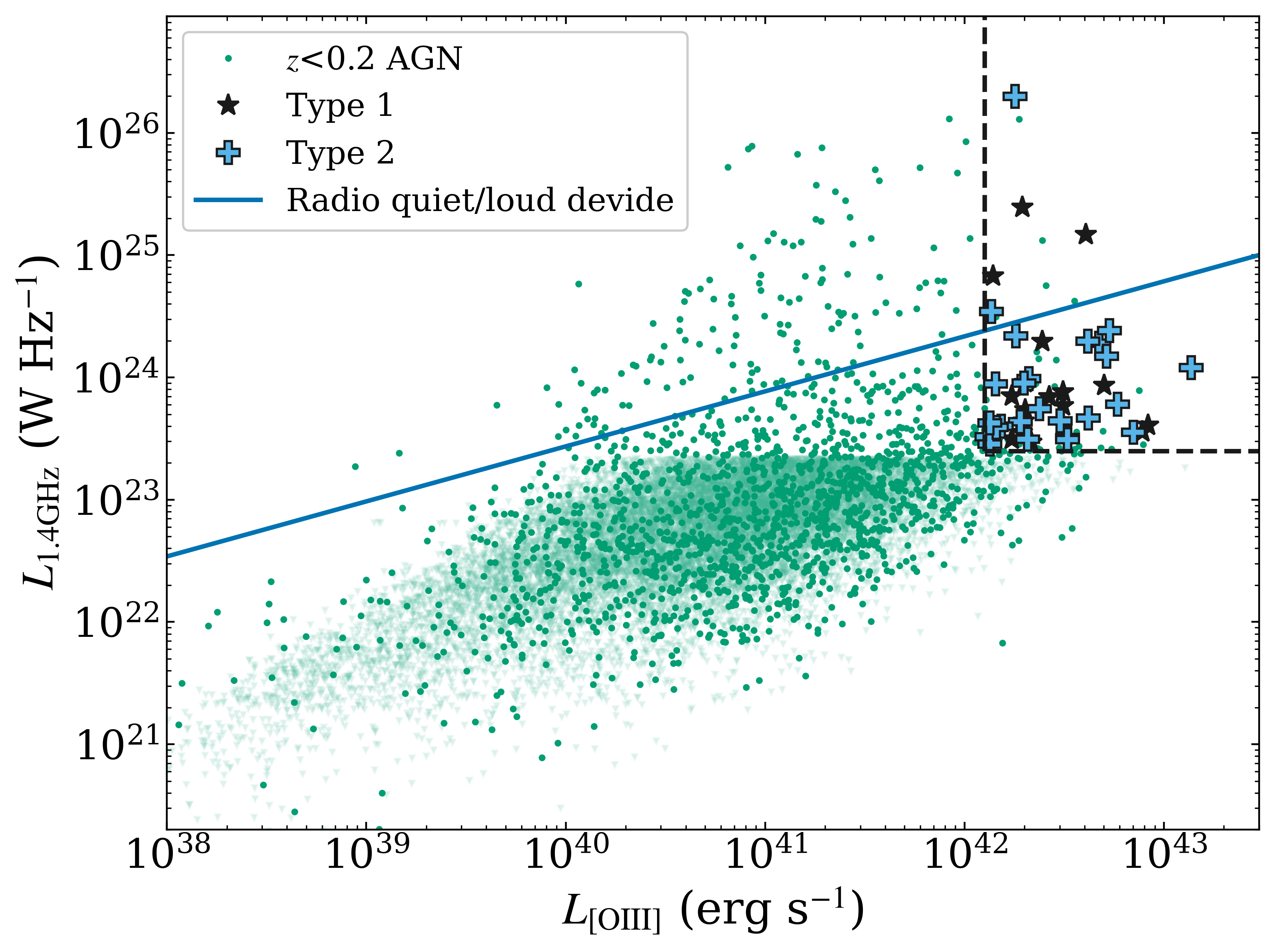}}
\caption{Total 1.4\,GHz luminosity against [O~\textsc{iii}] for our sample of type 1 (star symbols) and type 2 (plus symbols) quasars. The dashed lines show the primary criteria used to select the sample from the parent sample of $z<0.2$ AGN from \citet{Mullaney2013} (circles, with upper limits shown as pale green triangles). The solid line marks the divide between `radio-loud' and `radio-quiet' following \citet{Xu1999}. }

\label{fig:QFSsample}
\end{figure}

A detailed sample description has already been presented in \cite{Jarvis2021} and so we only provide a brief description of the sample selection here. The 42 quasars in the Quasar Feedback Survey were selected from the Sloan Digital Sky Survey (SDSS) parent sample of $\sim 24 000\,\, z < 0.4$, spectroscopically identified AGN presented in \cite{Mullaney2013}. Our sample selection was restricted to $z < 0.2$ with the highest AGN luminosities, using $L_\mathrm{[O III]} > 10^{42.11}\,$erg\,s$^{-1}$ and chosen to be dominated by moderate radio luminosities, using a cut of $L_\mathrm{1.4\,GHz} > 10^{23.45}\,$W\,Hz$^{-1}$. The radio luminosities are based on the flux densities from the NRAO VLA Sky Survey \citep[NVSS][]{Condon1998}, which were matched in \cite{Mullaney2013}, and assuming a uniform spectral index of $\alpha=-0.7$. The final sample of 42 quasars, highlighted as stars and crosses in Figure~\ref{fig:QFSsample}, was selected based on sky positions suitable for VLA scheduling, with right ascensions of RA=$10^{\circ} \-- 300^{\circ}$ and a declinations of $\mathrm{Dec}<24^{\circ}$ or $\mathrm{Dec}>44^{\circ}$. 

The sample is predominantly made up of sources classified as `radio-quiet' (37/42), based on the \textsc{[O III]} luminosity and the radio luminosity division of \cite{Xu1999} (see Figure~\ref{fig:QFSsample} and column (6) in  Table~\ref{Tab:sources1}), with 37/42 ($88\,$per\,cent) radio quiet sources and the rest radio loud. Furthermore, all but three of the targets have $L_\mathrm{1.4\,GHz} < 10^{25}$\,W\,Hz$^{-1}$. Based on the optical spectra, 25/42 sources are type 2 AGN and 17/42 are type 1 AGN.  

\subsection{\emerlin{} 6\,GHz observations and imaging}\label{sec:eMerlinData}
Observations of the 42 quasars from \cite{Jarvis2021} were carried out with \emerlin{} (programme ID: CY11205), excluding the Lovell telescope, between 2019 June and 2022 March across 26 observation runs for a total of $\sim 410\,\mathrm{hrs}$. The total on-source integration time was $\sim 350\,\mathrm{h}$ for the 42 targets, averaging to $\sim 8\,$hr per target. However, the total integration time per target was dependent on the minimum requested sensitivity, and therefore total integration time ranged from $4$ to $12\,$hrs per target. The observation setup frequency was 6550\,--\,7060\,MHz  centred at $\sim$ 6800\,MHz with at total bandwidth of 512\,MHz across 4 spectral windows (spws; at 128\,MHz/spws) The data was recorded at an aggregate bit rate of 1024\,Mbps with an integration time of $\mathrm{4\,s}$ and frequency resolution of 1.0\,MHz per channel. The flux density calibrator, 3C286, was observed for $\sim $40\,mins across each epoch. The 42 targets were observed for on-source scan length of 6\,mins per scan and the complex antenna gains tracked using a phase reference calibrator (for each target) for 2\,mins per scan, alternating between the target and the phase calibrator. For the bandpass calibrator, the source OQ 208 ($1407+2827$) was observed for $\sim 30\,$mins and was revisited every $4\,$hrs. The source OQ 208 was also used as the pointing calibrator in $\sim 1/3$ of the observation runs, otherwise the source 3C 84 (0319+4130) was used as the pointing and test calibrator.  

The raw uv-data was imported from the \emerlin{} archives and processed using the \emerlin{} \textsc{casa} pipeline. This is a Python based \textsc{casa} (v5.4+) package with standardized and optimized calibration parameters unique to \emerlin{} data\footnote{See \href{https://github.com/e-merlin/eMERLIN_CASA_pipeline}{https://github.com/e-merlin/eMERLIN\_CASA\_pipeline} for details.}. Before calibration, any bad data including Radio Frequency Interference (RFI) was auto-removed by the pipeline. The source 1331+305 (3C286) was used as a flux calibrator across all the observation runs since it has a well known source structure \citep[][]{PerleyButler2017}. The pipeline uses the structural input model for 3C286 which is adjusted for the total flux density scaled to account for the shortest available projected \emerlin{} baseline in the observations. This flux scaling was then applied to the bandpass and pointing-test calibrators. Assuming little-to-no time variability, bandpass calibration only corrected for the variations in the channel sensitivity across each spectral window. The complex gain solutions derived from the bandpass calibrator were applied on the flux calibrator and the pointing-test calibrators. All the derived solutions, flux density, bandpass and gain calibration were then applied across all the phase calibrators and targets in each run. Before imaging, any residual bad data and phase errors in the pipeline products were manually removed from the data. 

The targets were imaged using the \textsc{casa} TCLEAN task, where $2000 \times 2000$ maps with a cell size of $0.01\,$ arcsecs per pixel were created for each target. To reduce phase and amplitude noise, between one to three rounds of self-calibration were performed on sources with flux density $>1\,$mJy. 

\begin{figure*}
\centerline{\includegraphics[width=1\linewidth]{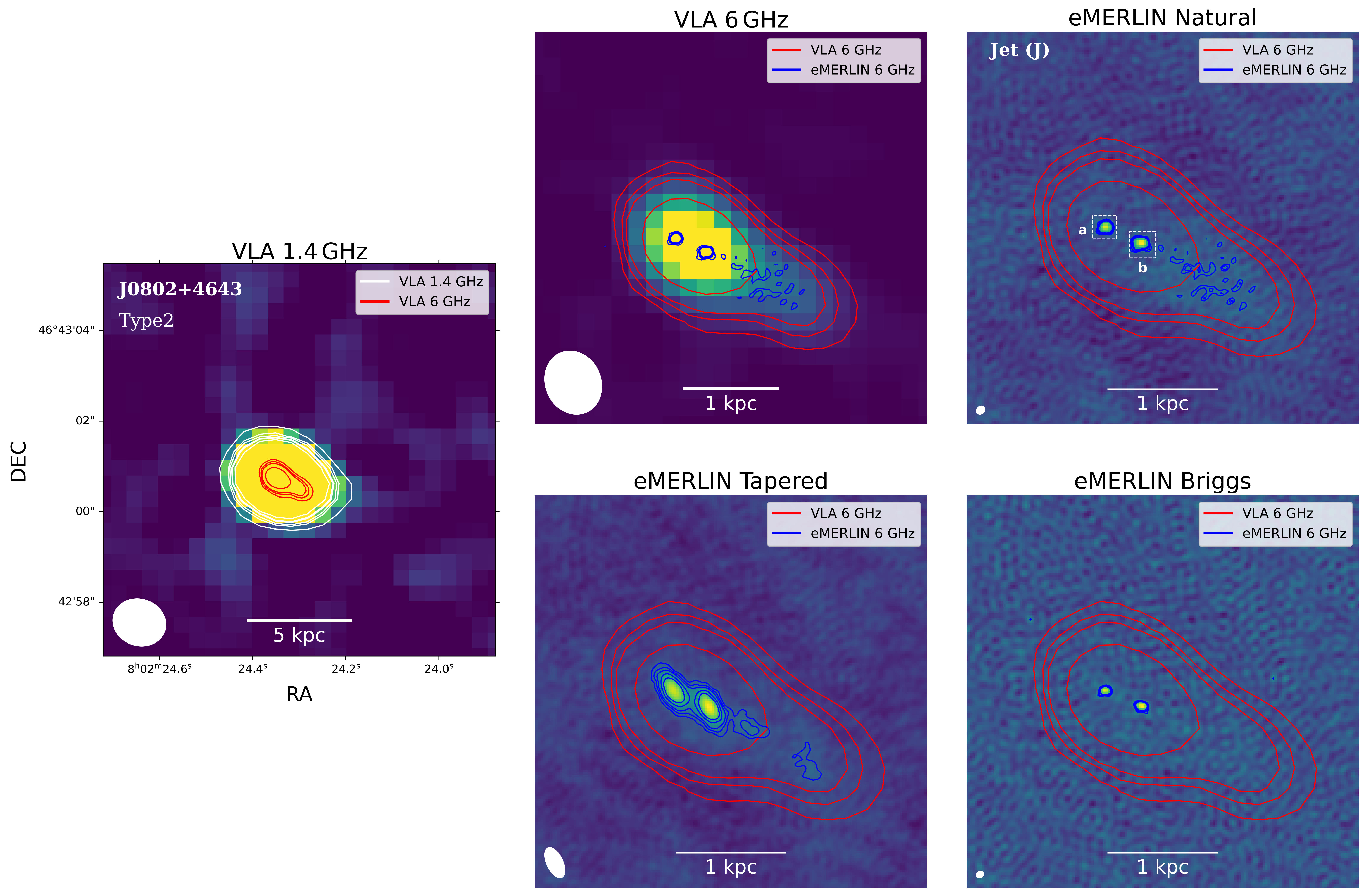}}
\caption[]{Example target to highlight the set of radio maps used in this work. \textit{Left:} VLA 1.4\,GHz image (white contours, with levels $1\sigma \times [5,10,12,15,16,20]$) overlaid with the VLA 6\,GHz image (red contours, with levels $1\sigma \times [4,8,12,16,32]$; in this and all other panels). \textit{Top middle:} the 6\,GHz VLA image overlaid with the \emerlin{} natural-weighted image (blue contours, with levels $1\sigma \times [4,5,6,7,8...]$; in this and all other panels). \textit{Top right:} \emerlin{} natural-weighted image (blue contours). Labelled regions on the \emerlin{} natural-weighted image are for any identified individual compact components (when more than one). \textit{Bottom middle:} \emerlin{} natural-weighted image with tapering applied (blue contours, with levels $1\sigma \times [4,5,6,7,8...]$). \textit{Bottom right:} \emerlin{} Briggs weighted image (blue contours, with levels $1\sigma \times [4,5,6,7,8...]$). Ellipses represent the synthesised beam for the corresponding background image in each panel. Scale bars are also shown to highlight physical size scales. Equivalent figures for all sources are provided in the Supplementary Material. }
\label{fig:J0802+4643}
\end{figure*}

Three weighting schemes were applied in the imaging: (1) natural weighting; (2) Briggs weighting with robust parameter r = 0.5; and (3) natural-weighted with \textit{uv}-tapering applied at $4\,\mathrm{M}\lambda$. Applying \textit{uv}-tapering to the data for the latter maps, helps to probe the more extended emission. Overall, the naturally weighted maps provide the maximum point source sensitivity, the Briggs weighted maps provide the highest spatial resolution, while the \textit{uv}-tapered maps produce the maximum sensitivity for extended emission. The natural weighted maps have a typical rms of $\sim 20\,\microJybm$, and a typical restoring beam $\sim 50 \times 30$\,mas. The Briggs weighting provides the highest resolution with a restoring beam of $\sim$ $30 \times 10\,$mas and a typical central rms of $\sim  50\,\microJybm$, while the tapered maps provide a restoring beam size of $\sim 100 \times 80\,$mas and a lower point source sensitivity with a typical central rms of $\sim 60\,\microJybm$. The rms noise and restoring beam sizes/shapes vary across the sources depending on the exact exposure times, uv coverage, and level of self-calibration. All the \emerlin{} maps we produced (i.e., 3 maps for each of the 42 targets) are released with this work (see Data Availability section). Figure~\ref{fig:J0802+4643}, shows a set of the three \emerlin{} maps that were produced for one example target, with the equivalent figures for the rest of the targets provided in the Supplementary Material\footnote{These are also available here: \url{https://doi.org/10.25405/data.ncl.27604872}}. 

As part of our discussion in this work, and to help with the identification/verification of features identified in the \emerlin{} maps, we make use of the VLA maps presented in \cite{Jarvis2021} of the same targets. Briefly, we utilise the L-band ($\sim$1.4\,GHz) and C-band ($\sim$6\,GHz) maps, both taken with the A-array configuration of the VLA, and imaged with Briggs weighting (robust parameter r=0.5). The corresponding synthesized beams have major axis sizes of $\sim$1\,arcsec and $\sim$0.3\,arcsec, respectively. These VLA maps for an example target are shown in Figure~\ref{fig:J0802+4643}, with the equivalent figures for the whole sample presented in the Supplementary Material.

\subsection{Image analysis}\label{sec:images}
The \textsc{casa} task IMFIT was used for source fitting any components that were identified in the radio maps with peak signal-to-noise ratios SNR$\gtrsim$4$\sigma$, and which could be well described by a two-dimensional Gaussian model. Where multiple components were identified, their properties were calculated separately. These are highlighted in the figures (see example in Figure~\ref{fig:J0802+4643}) and are listed separately in Table~\ref{Tab:sources1}.  Any identified diffuse structures (not characterised by the fitting or properties in Table~\ref{Tab:sources1}) form part of the morphology discussion in Sections~\ref{sec:morphology} and Section~\ref{sec:radio-jets}. We note that the identified \emerlin{} structures are always co-located with known existing 6\,GHz emission as seen in the VLA maps, providing further confidence in their reliability. 

The natural-weighted maps were primarily used for the extraction of the radio properties with \textsc{casa}, as these maps provide the maximum sensitivity to compact structures. However, in 5 sources, structures were only identified in the tapered maps, and these were used instead (see Section~\ref{sec:analysis}). These targets are highlighted in Table~\ref{Tab:sources1}. The \textsc{casa} IMFIT task fits two-dimensional elliptical Gaussians to an intensity distribution region that has been manually-selected on a radio map. The fitting process returned the source coordinates at the radio peak, the source size deconvolved from the beam (the angular full width at half-maximum, FWHM, at both the major and minor axes), the position angles (PA; in degrees), the integrated flux density and the peak brightness. In some instances, during the fitting process, some sources remained undeconvolved from the beam (unresolved, with no size constraints) or were only marginally resolved along one or both directions (upper limits returned on the size). These results, the associated errors from the fits, and the derived radio properties are all provided in Table~\ref{Tab:sources1}. 

A visual inspection of the radio maps, combined with the measured radio properties, and derived properties (such as the brightness temperature measurements) are used to investigate the origin of radio emission in this sample of quasars at $\sim$sub-kiloparsec scales, as presented in Section~\ref{sec:analysis} and discussed in Section~\ref{sec:discussions}. 

\section{Results}\label{sec:analysis}
Three sets of radio maps from our 6\,GHz \emerlin{} data were produced for each of the 42 targets of the Quasar Feedback Survey (see example in Figure~\ref{fig:J0802+4643}). As described in Section~\ref{sec:images}, these maps cover a range of higher spatial resolution (Briggs weighting), optimum sensitivity to compact structures (natural weighting) and sensitivity to more extended structures (natural weighting with tapering). The natural-weighted radio maps were primarily those used for the extraction of the compact radio structure properties. However, we find that 10/42 sources remained undetected in these maps, but with 5/10 of these sources being detected in the tapered maps. In these 5 cases, the radio properties were extracted from the tapered maps, and are labelled with an asterisk in Table~\ref{Tab:sources1}. In 5 sources, more than one compact structure was identified in the maps. For these sources, the properties of the individual structures were separately extracted, labelled with `a', `b', etc, (see example in Figure~\ref{fig:J0802+4643}), and are highlighted in bold in Table~\ref{Tab:sources1}. 

In Section~\ref{sec:morphology}, we present our source classification based on a visual inspection of the radio morphology. The flux density measurements across sub-kiloparsec to kiloparsec scales are presented in Section~\ref{sec:fluxes}, with the radio luminosity measurements at 6\,GHz and brightness temperature presented in Sections~\ref{sec:luminosity} and Section \ref{sec:TB}, respectively.

Table~\ref{Tab:sources1} presents the properties obtained for the radio structures identified in the \emerlin{} images, in addition to some derived quantities (described later in this section), and other relevant source properties used for our results and discussion. The columns are as follows:\\
Column (1) gives the source ID in international coordinates. \\
Column (2) and (3) give the source positions from \emerlin{} in J2000. A `U' indicates that the source was non-detected in \emerlin{} and therefore all the \emerlin{} radio measurements remained undetermined, with a corresponding `U' in all relevant columns.\\
Column (4) gives the redshift values from SDSS (DR7). \\
Column (5) is the projected linear size (the deconvolved major axis) of the source. The 7 sources italicised with double asterisks remained completely undeconvolved based on the fitting results, and conservative upper limits are provided, using the size of the restoring beam.  \\
Column (6) states the type of AGN classification whether type 1 or 2 (based on optical spectra) and if radio-loud (RL) or radio-quiet (RQ) based on \cite{Xu1999}. See Figure~\ref{fig:QFSsample}. \\
Column (7) gives the total flux density (mJy) measurements. \\
Column (8) gives the peak brightness (mJy/beam) measurements. \\ 
Column (9) provides the derived brightness temperature ($T_\mathrm{B}\,$)K measurements. The sources italicised with double asterisks remained undeconvolved from the beam and the lower limits on the $T_\mathrm{B}\,$ are derived based on the size of the restoring beam. \\
Column (10) list the derived radio luminosities of the extracted radio structures at 6\,GHz ($L_\mathrm{6\,GHz}$).\\
Column (11) provides the spectral indices ($\alpha$, over 4--8\,GHz) of the core emission, as obtained from the $\sim$0.3\,arcsecond resolution C-band VLA images (derived in \citealt{Jarvis2021}). \\
Column (12) lists our radio morphology classification, based on visual inspection only.\\
Column (13) lists our final classification of the origin of the radio emission. Question marks (?) indicates an unknown classification, while J+21 are sources classified as a radio-AGN in \cite{Jarvis2021}, but not in this work.\\

\begin{landscape}
\begin{table}
    \caption{The radio properties for the 42 Quasar Feedback Survey targets from \emerlin{} at $6\,$GHz.}
    \label{Tab:sources1}
    \small
    \centering
    \begin{tabular}{lccccccccccccc}
      
        \hline
        Source ID & RA(J2000) & DEC(J2000) & $\mathrm{z}$ & Linear & Type & Flux & Peak &$T_\mathrm{B}$ & $L_\mathrm{6\,GHz}$ & $\alpha_\mathrm{core}$ & Radio & Our \\
        
         & & & & size (pc) &  & ($\mathrm{mJy}$) & ($\mathrm{mJy\,beam^{-1}}$) & (K) & W\,Hz$^{-1}$ & Jarvis+21 & Morphology & class \\

         (1) & (2) & (3) & (4) & (5) & (6) & (7) & (8) & (9) & (10) & (11) & (12) & (13) \\
        \hline
        
        J0749+4510 & 07:49:06.5077 & +45:10:33.8758 & 0.192 & \textit{<603.47$^{**}$} & $1,$RL & 38.66\plus{0.65} & 39.45\plus{0.28} & \textit{>1.6$\times$10$^{5**}$} & $3.22\times10^{24}$ &  0.31\plus{0.07} & Compact & radio-AGN \\
        
        J0752+1935 & 07:52:17.8557 & +19:35:42.1305 & 0.117 & 78.88\plus{17.7} & 1,RQ & 0.734\plus{0.049} & 0.48\plus{0.02} & $1.9 \times 10^4$ &$2.66\times10^{22}$ & -1.2\plus{0.1} & Jet & radio-AGN \\
        
        J0759+5050 & 07:59:40.9488 & +50:50:23.9955 & 0.055 & 135.59\plus{13.2} & 2,RQ & 5.61\plus{0.31} & 3.83\plus{0.13} & $2.5 \times 10^4$ & $4.09\times10^{22}$ & -1.22\plus{0.06} & Jet & radio-AGN \\
        
        \textbf{\textit{J0802+4643a}} & 08:02:24.3563  & +46:43:00.7791  & 0.121 & 92.46\plus{13.9} & 2,RQ & 0.831\plus{0.077} &	0.403\plus{0.027} & $1.0 \times 10^{4}$ & $3.22\times10^{22}$ &  -1.135\plus{0.002} & \textbf{\textit{Jet}} &  \textbf{\textit{radio-AGN}} \\

        \textbf{\textit{J0802+4643b}} & 08:02:24.3420  & 46:43:00.7116  & 0.121 & 108.21\plus{14.4} & 2,RQ & 1.009\plus{0.1} & 0.431\plus{0.031}& $8.7 \times 10^{3}$ & $3.91\times10^{22}$ &  &  &  \\
        
        J0842+0759 & 08:42:05.5753 & +07:59:25.5448 & 0.134 & \textit{<171.84$^{**}$} & $1,$RQ & 1.18\plus{0.099} & 1.233\plus{0.005} & \textit{>1.9$\times$10$^{4**}$} & $5.20\times10^{22}$ & -0.4\plus{0.3} & Jet & radio-AGN \\
        
        J0842+2048 & 08:42:07.5143 & +20:48:40.1529 & 0.181 & $<173.06$ & $1,$RQ & 1.003\plus{0.069} & 1.0\plus{0.035} & $>7.3 \times 10^4$ & $8.78\times10^{22}$ & -0.7\plus{0.02} & Compact & radio-AGN\\
        
        J0907+4620 & 09:07:22.3459 & +46:20:18.1822 & 0.167 & 42.75\plus{2.9} & 2,RL & 7.03\plus{0.097} & 6.124\plus{0.046} & $1.3 \times 10^6$ & $5.09\times10^{23}$ &  -0.6\plus{0.1} & Jet & radio-AGN \\
        
        J0909+1052 & 09:09:35.5039 & +10:52:10.5827 & 0.166 & $<152.57$ & $2,$RQ & 0.66\plus{0.079} & 0.505\plus{0.035} & $>2.0 \times 10^4$ & $5.11\times10^{22}$ & -1.12\plus{0.06} & Compact & ? \\
        
        J0945+1737 & 09:45:21.3389 & +17:37:53.2662 & 0.128 & 61.86\plus{13.9} & 2,RQ & 3.76\plus{0.14} & 2.871\plus{0.062} & $2.0 \times 10^5$ & $1.60\times10^{23}$ &  -0.89\plus{0.07} & Irregular & radio-AGN \\
        
        J0946+1319 & 09:46:52.5769 & +13:19:53.8710 & 0.133 & 151.24\plus{31.7} & 1,RQ & 1.27\plus{0.14} & 0.422\plus{0.035} & $3.7 \times 10^3$ & $5.95\times10^{22}$  & -1.0\plus{0.3} & Jet  & radio-AGN \\
        
        \textbf{\textit{J0958+1439a}} & 09:58:16.8926 & +14:39:23.9573 & 0.109 & 217.37\plus{29.6} & 2,RQ & 0.664\plus{0.064} &	0.159\plus{0.013}  & $8.1 \times 10^{2}$ & $2.07\times10^{22}$ &  -1.23\plus{0.01} & \textbf{\textit{Jet}} & \textbf{\textit{radio-AGN}} \\

        \textbf{\textit{J0958+1439b}} & 09:58:16.8943 & +14:39:23.8242 & 0.109 & 119.16\plus{28.8} & 2,RQ & 0.476\plus{0.046} &	0.186\plus{0.013} & $2.1 \times 10^3{}$ & $1.49\times10^{22}$ &  & &  \\

        \textbf{\textit{J0958+1439c}} & 09:58:16.8988 & +14:39:23.7288 & 0.109 & 162.72\plus{33.3} & 2,RQ & 0.356\plus{0.042} &	0.145\plus{0.012} & $1.7 \times 10^{3}$ & $1.11\times10^{22}$ &  & &  \\

        \textbf{\textit{J0958+1439d}} & 09:58:16.9039 & +14:39:23.4930 & 0.109 & 83.42\plus{19.6} & 2,RQ & 0.504\plus{0.038} &	0.281\plus{0.014} & $6.4 \times 10^{3}$ & $1.57\times10^{22}$ &  & & \\
        
        J1000+1242 & 10:00:13.1504 & +12:42:26.2654 & 0.148 & 49.35\plus{5.3} &	2,RQ & 7.087 \plus{0.051} & 6.507\plus{0.024} & $1.2 \times 10^6$ & $4.04\times10^{23}$ & -0.7\plus{0.05} & Compact & radio-AGN \\
        
        J1010+0612 & 10:10:43.3638 & +06:12:01.3871 & 0.098 & 37.71\plus{4.9} & 2,RQ & 18.55\plus{0.13} & 17.542\plus{0.058} & $3.8 \times 10^6$ & $4.55\times10^{23}$ & -1.11\plus{0.02} & Compact & radio-AGN \\

        J1010+1413 & 10:10:22.9563 & +14:13:00.8012 & 0.199 & \textit{<257.78$^{**}$} & 2,RQ  & 2.197\plus{0.083} & 2.026\plus{0.042} & \textit{>2.8$\times$10$^{4**}$} & $2.45\times10^{23}$ & -0.9\plus{0.04} & Compact & J+21 \\
        
        J1016+0028 & U & U  & 0.116 & U & 2,RQ & U & U & U & U & U &  $\mathrm{Non-detected}$ & J+21 \\
        
        J1016+5358 &10:16:23.7731 & +53:58:06.0736 & 0.182 & $<410.85$ & 2,RQ & 0.183\plus{0.051} & 0.146\plus{0.023} & $>6.3 \times 10^2$ & $1.79\times10^{22}$  & -1.3\plus{0.1} & Jet & radio-AGN \\
        
        J1045+0843 & 10:45:05.1604 & +08:43:39.1434 & 0.125 & \textit{<532.10$^{**}$} & 1,RQ & 0.995\plus{0.095} & 0.829\plus{0.029} & \textit{>3.7$\times$10$^{3**}$} & $4.10\times10^{22}$ & -1.05\plus{0.05} & Compact & ? \\
        
        J1055+1102$^{*}$ & 10:55:55.2787 & 11:02:51.7446 & 0.145 & 406.66\plus{86.6} & 2,RQ & 0.338\plus{0.038} & 0.114\plus{0.09} & $5.4 \times 10^{2}$ & $1.94\times 10^{22}$ &  -1.11\plus{0.09} & Jet & radio-AGN \\

        J1100+0846 & 11:00:12.3843 & +08:46:16.3234 & 0.1 & 173.31\plus{26.7} & 2,RQ & 9.8\plus{0.58}	& 6.78\plus{0.25} & $4.7 \times 10^4$ & $2.49\times10^{23}$ & -1.04\plus{0.04} & Compact & J+21 \\

        \textbf{\textit{J1108+0659a$^{*}$}} & 11:08:51.0344 & 06:59:01.2629 & 0.181 & 544.34\plus{75.5} & 2,RQ & 0.189\plus{0.029} & 0.0782\plus{0.009} & $6.4 \times 10^{2}$ & $1.86 \times 10^{22}$ & -1.4\plus{0.1} & \textbf{\textit{Jet}} & \textbf{\textit{radio-AGN}} \\
        
        \textbf{\textit{J1108+0659b$^{*}$}} & 11:08:51.0614 & 06:59:00.9664 & 0.181 & 380.73\plus{81.8} & 2,RQ & 0.158\plus{0.017} & 0.102\plus{0.067} & $1.1 \times 10^{3}$ & $1.55 \times 10^{22}$ & &  & \\

        J1114+1939 & U & U & 0.199 & U & 2,RQ & U & U & U & U & -0.83 & Non-detected & ? \\
        
        J1116+2200 & U & U & 0.143 & U & 2,RQ & U & U & U & U & -1.22\plus{0.05} & Non-detected & ?  \\

        J1222-0007 & 12:22:17.8631 & -00:07:43.6912 & 0.173 & 527.93\plus{200.2} & 2,RQ &  0.384\plus{0.073} & 0.113\plus{0.016} & $4.0 \times 10^2$ & $3.13\times10^{22}$ & -0.87\plus{0.1} & Jet & radio-AGN \\
  
        J1223+5409 & 12:23:13.21721 & 54.09.06.50966 & 0.156 & 109.85\plus{27.3} & 1,RL & 12.18\plus{0.62} & 9.59\plus{0.27} & $4.5 \times 10^{5}$ & $7.51\times10^{23}$ & -0.5\plus{0.3} & Compact & radio-AGN \\
        
        J1227+0419 & 12:27:39.8254 & +04:19:32.3342 & 0.18 & \textit{<767.48$^{**}$} & 1,RQ & 1.911\plus{0.071} & 2.005\plus{0.039} & \textit{>2.1$\times$10$^{3**}$} & $1.74\times10^{23}$ & -1.02\plus{0.04} & Compact & ? \\
        
        J1300+0355 & 13:00:07.9960 & +03:55:56.6288 & 0.184 & 78.43\plus{14.3} & 1,RQ & 17.481\plus{0.056} & 17.25\plus{0.028} & $3.5 \times 10^6$ & $1.36\times10^{24}$ & 0.2\plus{0.1} & Compact & radio-AGN \\
        
        J1302+1624 & 13:02:58.8441 & +16:24:27.8248 & 0.067 & 196.26\plus{35.8} & 1,RQ & 3.605\plus{0.062} & 2.908\plus{0.029} & $5.0 \times 10^{3}$ & $3.89\times10^{22}$ & -0.87\plus{0.07} & Jet & radio-AGN \\
        
        \textbf{\textit{J1316+1753a}} & 13:16:42.8993 & +17:53:32.6408 & 0.15 & \textit{<153.84$^{**}$} & 2,RQ & 0.35\plus{0.043} & 0.345\plus{0.024} & \textit{>6.8$\times$10$^{3**}$} & $2.18 \times 10^{22}$ & -1.18\plus{0.04} & \textbf{\textit{Jet}} &  \textbf{\textit{radio-AGN}} \\

        \textbf{\textit{J1316+1753b}} & 13:16:42.9218 & 17:53:32.4226 & 0.15 & 141.96\plus{42.9} & 2,RQ & 0.286\plus{0.043} &	0.136\plus{0.014} & $2.5 \times 10^{3}$ & $1.79 \times 10^{22}$ &  & & \\

        J1324+5849 & 13:24:18.2529 & +58:49:11.7115 & 0.192 & 31.66\plus{19.5} & 1,RQ & 0.788\plus{0.09} & 0.691\plus{0.047} & $4.7 \times 10^5$ & $8.18 \times 10^{22}$ & -0.94\plus{0.02} & Compact & radio-AGN \\
        
        J1347+1217 & 13:47:33.3611 & +12:17:24.2239 & 0.121 & 64.11\plus{14.4} & 2,RL & 1958\plus{0.012} & 1834\plus{0.0057} & $9.9 \times 10^7$ & $6.98 \times 10^{25}$ & -0.4\plus{0.2} & Compact & radio-AGN \\
        
        J1355+2046 & 13:55:50.1994 & +20:46:14.5014 & 0.196 & 522.86\plus{136.8} & 1,RQ & 0.398\plus{0.065} & 0.173\plus{0.017} & $2.6 \times 10^3$ & $4.31\times10^{22}$ & -0.92\plus{0.03} & Jet & radio-AGN \\
        
        J1356+1026 & 13:56:46.1064 & +10:26:09.0619 & 0.123 & 185.20\plus{7.3} & 2,RQ  & 12.65\plus{0.13} &  10.284\plus{0.045} & $5.2 \times 10^5$ & $5.05\times10^{23}$ & -1.09\plus{0.06} & Compact & radio-AGN \\

        J1430+1339 & 14:30:29.8716 & +13:39:11.8827 & 0.085 & 141.67\plus{37.9} & 2,RQ & 3.58\plus{0.33} & 3.03\plus{0.14} & $2.9 \times 10^4$ & $6.53\times10^{22}$ & -1.2\plus{0.1} & Jet & radio-AGN \\
        
        J1436+4928 & 14:36:07.20911 & +49:28:58.5309 & 0.128 & 73.19\plus{42.5} & 2,RQ & 0.331\plus{0.059} & 0.272\plus{0.028} & $1.3 \times 10^5$ & $1.44\times10^{22}$ & -1.1\plus{0.1} & Compact & radio-AGN \\

        J1454+0803 & 14:54:34.3504 & +08:03:36.6701 & 0.13 & 241.63\plus{105.3} & 1,RQ & 0.638\plus{0.109} & 0.332\plus{0.038} & $1.8 \times 10^3$ & $2.88 \times 10^{22}$ & -1.1\plus{0.1} & Jet & radio-AGN \\
        
        J1509+1757$^{*}$ & 15:09:13.7902 & 17:57:10.1417 & 0.171 & 324.61\plus{132.2} & 1,RQ & 0.287\plus{0.032} & 0.198\plus{0.013} & $2.0 \times 10^{3}$ & $2.48 \times 10^{22}$ & -1.4\plus{0.1} & Jet & radio-AGN \\
        
        J1518+1403 & U & U & 0.139 & U & 2,RQ & U & U & U & U & -1.2 & Non-detected & ?  \\
        
        J1553+4407 & U & U & 0.197 & U & 2,RQ & U & U & U & U & -0.2 & Non-detected & J+21 \\
        
        J1555+5403$^{*}$ & 15:55:01.4393 & 54:03:27.0033 & 0.18 & 383.11\plus{26.0} & 1,RQ & 0.41\plus{0.028} & 0.083\plus{0.005} & $4.5 \times 10^{2}$ & $3.73 \times 10^{22}$ & -1.0\plus{0.1} & Jet & radio-AGN \\
        
        \textbf{\textit{J1655+2146a$^{*}$}} & 16:55:51.3569 & 21:46:03.0907 & 0.154 & 540.67\plus{151.7} & 1,RQ & 0.39\plus{0.049} & 0.189\plus{0.016} & $3.6\times 10^{2}$ & $2.57 \times 10^{22}$ & -1.15\plus{0.04} & \textbf{\textit{Jet}} & \textbf{\textit{radio-AGN}}  \\

        \textbf{\textit{J1655+2146b$^{*}$}} & 16:55:51.4088  & 21:46:02.1131 & 0.154 & \textit{<551.70}$^{**}$ & 1,RQ & 0.149\plus{0.018} & 0.187\plus{0.011} & \textit{>3.5$\times$10$^{2**}$}  & $9.82 \times 10^{21}$ & & &   \\
        
        J1701+2226 & 17:01:58.2478 & +22:26:41.8308 & 0.197 & 251.04\plus{103.3} & 1,RL & 3.56\plus{0.15} & 3.22\plus{0.09} & $1.3 \times 10^5$ & $3.38\times10^{23}$ & -0.12\plus{0.02} & Jet & radio-AGN \\

        J1715+6008 & 17:15:44.0246 & +60:08:35.5583 & 0.157 & 213.02\plus{53.3} & 2,RQ & 1.26\plus{0.3}	& 0.334\plus{0.064} & $5.2 \times 10^3$ & $8.57\times10^{22}$ & -1.08\plus{0.08} & Jet & radio-AGN \\
         
        \hline

\end{tabular}
\end{table}
\end{landscape}

\subsection{Sub-kiloparsec radio morphology}\label{sec:morphology}
\begin{figure*}
\centerline{\includegraphics[width=0.85\linewidth]{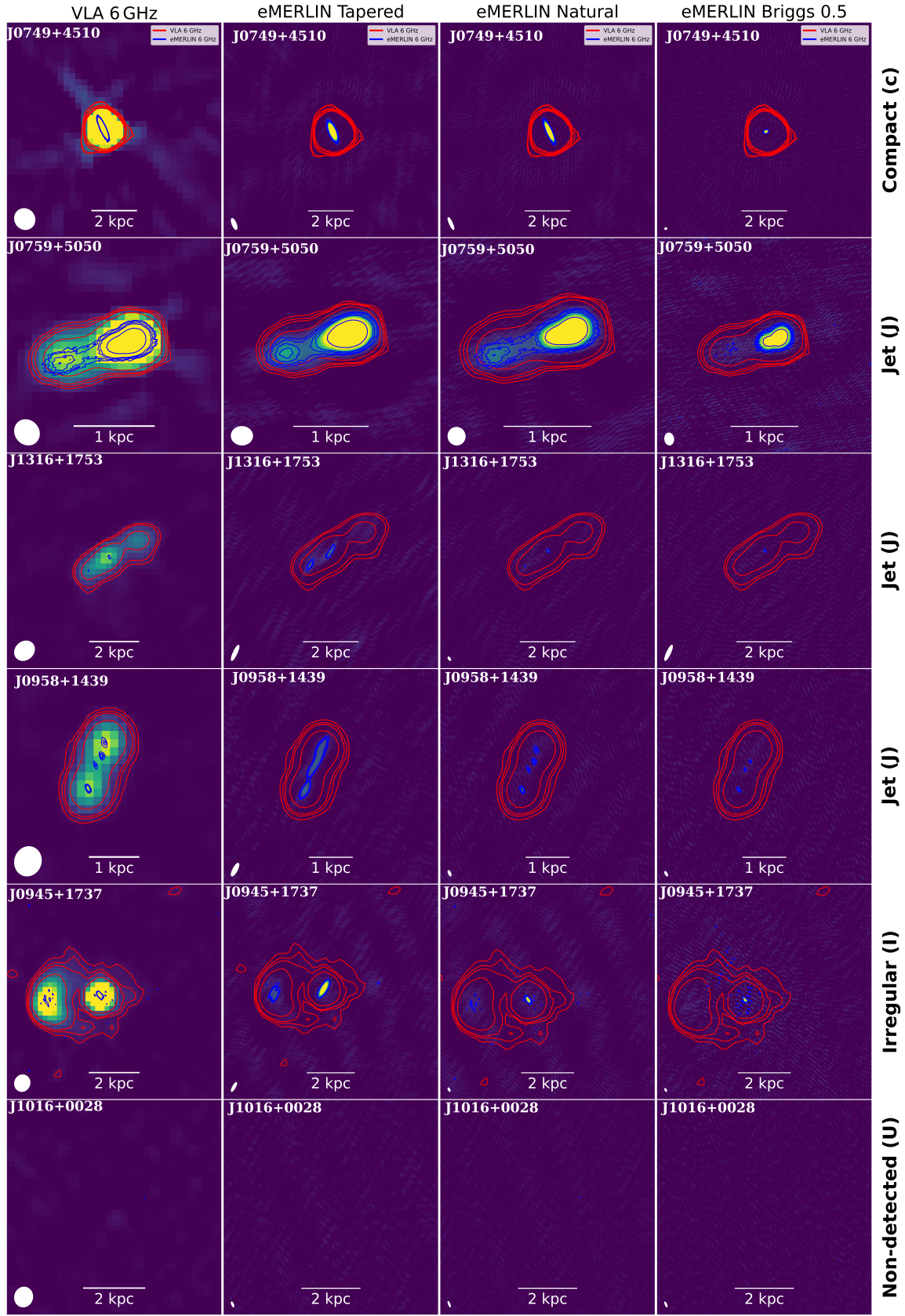}}
\caption{Example radio maps of 6 different targets (one in each row) to highlight the different radio morphology classifications. The first column shows VLA 6\,GHz radio maps, with the corresponding contours shown in red in all other panels. The remaining panels show the three different \emerlin{} maps made for each source, with corresponding contours in blue (tapered, natural weighting and Briggs weighting). The morphology classification for each target (compact, jet, irregular or non detected) is shown on the right. } 
\label{fig:sample}
\end{figure*}

Faint radio galaxies show a diverse range of morphologies at galactic scales from unresolved compact radio cores, sub-kiloparsec radio jets, side lobes, hotspots and diffuse emission  \citep[e.g.][]{Baldi2019,Pierce2020,Hardcastle2020,Saikia2022,Williams2023}. Probing the radio morphology of these quasars is paramount and particularly useful in understanding AGN feedback processes, and the corresponding implication on galaxy evolution. Therefore, using the presence (or lack) of extended emission, we describe the radio morphology of our target sample based on the four categories described below. Examples from the different categories are shown in Figure~\ref{fig:sample}. This approach is a simplified approach to that used in similar previous work \cite[e.g.,][]{Baldi2018,Jarvis2021}. Based on visual inspection only, we classified our sources into:

\begin{itemize}
    \item Compact (C): targets that only visibly show compact emission, i.e., without any signs of extended, or multiple, radio structures. We note that these sources may still be extended when deconvolved from the beam during the fitting process (Section~\ref{sec:images}).\\
    
    \item Jet (J): targets that show cores with one-sided or two-sided jet-like features (i.e., clearly asymmetric/marginally resolved and collimated in one or two directions from the core). Additionally, targets that show multiple, but well collimated, radio knots were also classified as jet-like since well collimated symmetrical double/triple radio knots are indicative of radio-AGN jets that are terminated in the dense ISM \citep[e.g.][]{Leipski2006,Middelberg2007,Odea2021}. This classification combines multiple sub-classifications of jet-like structures used in other works \cite[e.g.,][]{Baldi2018,Jarvis2021}. We note that this definition is describing the morphology. It may be physically associated with an AGN-driven jet, or potentially a collimated wind-driven outflow \citep[see discussion in][]{Harrison2024}.\\

     \item Irregular (I): any source that is visibly extended but does not fit the jet criteria, was classified as irregular. \\
     
     \item Non-detected (U): classification for any sources that remain undetected across all the \emerlin{} maps (natural-weighted, tapered, and Briggs weighting).

\end{itemize}

Visual inspection for the classification of the radio morphologies was primarily performed using the natural-weighted maps; however, the tapered maps often provided confirmation of low surface brightness extended structures seen (e.g., see J1316+1753 and J0945+1747, in Figure~\ref{fig:sample}). For the 5/10 sources that were only detected in the tapered maps, the tapered maps were used for morphology classification. 

Based on this visual inspection, 15/42 targets were classified as compact sources, 21/42 as jet-like, 1/42 as irregular and 5/42 as non-detected. Overall, the natural-weighted maps showed mainly compact structures, with isolated and well collimated radio knots (classified as multi-component sources; e.g., J0802+4643 in Figure~\ref{fig:J0802+4643}). In some cases, one-sided jet-like structures were clearly identified (e.g., J0759+5050 in Figure~\ref{fig:sample}). The Briggs-weighting maps further isolated/confirmed the compact cores in most targets (e.g., J0945+1747 in Figure~\ref{fig:sample}) and the well-collimated nature of the radio knots/structures. On the other hand, the tapered maps often drew out some low surface brightness extended emission, which sometimes connected the hotspots observed in the higher resolution maps (e.g., J0958+1439 in Figure~\ref{fig:sample}).

\subsection{Sizes and flux densities of sub-kiloparsec radio structures}\label{sec:fluxes}

\begin{figure}
\centerline{\includegraphics[width=1\linewidth]{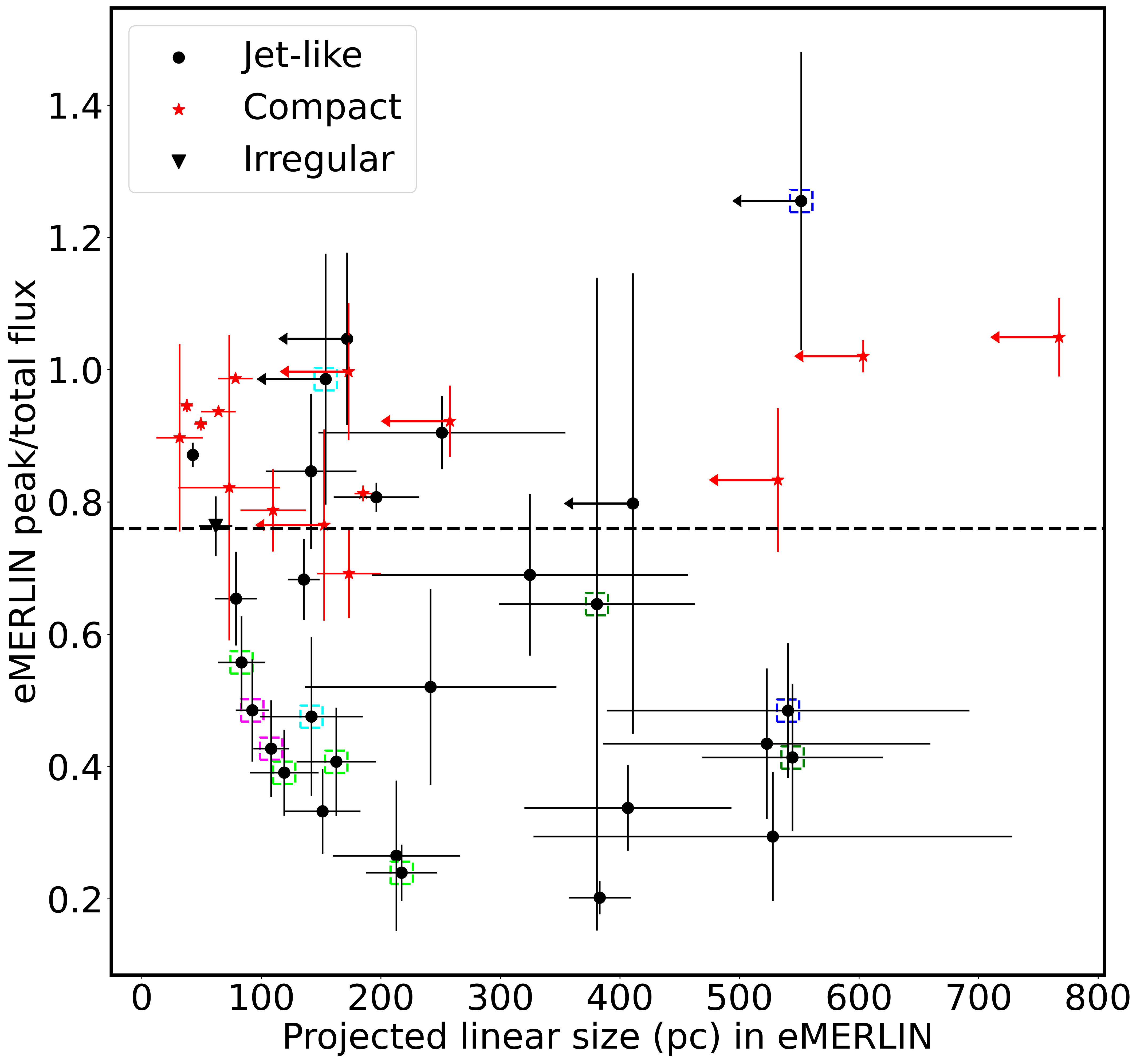}}
\caption[]{The ratios ($R$) of the peak brightness to integrated flux density against the projected linear size for all the structures identified in the \emerlin{} images. The multi-component sources are highlighted with dashed squares, using different colours for different sources. The median peak brightness to integrated flux density ratio is $R = 0.76$ (dashed line). As expected, $\sim 80\,$per cent of the sources showing jet-like and extended morphology lie below this median, while $\sim 93\,$per cent of the compact sources lie above this median, indicating that a large fraction of the flux density in the compact sources can be attributed to the central core.}
\label{fig:flux}
\end{figure}

Figure~\ref{fig:flux} shows the ratios ($R$) of the peak brightness to the integrated flux density as a function of the projected linear size in \emerlin{}, for each of the structures identified and fit (see Section~\ref{sec:images}). Across the 37 detected targets, there are 44 detected structures in total (i.e., due to the 2--4 multiple components observed in 5 of the targets; see Table~\ref{Tab:sources1}). Excluding all the upper limits, the projected linear sizes range from $\sim$30--540\,pc, with a median size of $\sim147$\,pc, Of the 37 \emerlin{} detected targets, 8 only have upper limits on the projected linear sizes for any of the structures identified. In all these 8 targets, only a single compact component is identified, indicating the radio emission on these scales is dominated by a compact, unresolved core. However, the upper limits are not very constraining in most cases due to beam sizes (see Section~\ref{sec:images}), reaching upper limit values of $\lesssim$770\,pc. 

Across all 44 detected structures, the median peak/total flux density ratio is $R = 0.76$. This indicates that, on average, only $\sim 20\,$per\,cent of the integrated flux density could be attributed to more extended radio emission in the structures, and they are typically very compact, confirming what was measured for the projected linear sizes. As can be seen in Figure~\ref{fig:flux}, this peak/total flux density ratio ranges from $\sim R=0.2 \--1.0$, across all the structures. Overall, the sources classified as compact based on their morphology (Section~\ref{sec:morphology}) have high peak/total flux density values, with $\sim 93\,$per\,cent (14/15) above the median value. This is true for all of the unresolved sources, further indicating that they are point sources and that nearly all of the flux can be attributed to the central core. We note that all the four sources above $R>1$ (J0749+4510, J0842+0759, J1227+0419 and J1655+2146b), have error bars consistent with R$\sim$1.0, and remain undeconvolved from the beam, further indicating that they are dominated by central, unresolved, compact cores. This is true even for the two sources with the least constraining upper limits on the projected linear sizes with upper limits of 603 and 767\,pc (i.e., J0749+4510 and J1227+0419).

We investigated the 9 jet-like or irregular targets, that showed a high peak/total flux density ratio $R>0.76$ (i.e., greater than the median). These targets are classified as jet-like based on visual inspection due to the well collimated extended emission structure and/or the presence of multiple radio knots in the systems seen across the whole target. These extended structures are not captured by the fitting of the \textit{individual} components which are presented in Figure~\ref{fig:flux} and Table~\ref{Tab:sources1}, which may be highly compact in themselves. We therefore conclude that for the extended or jet-like sources showing $R>0.76$, this can be attributed to our fitting under sampling the \textit{total} integrated flux of the source.  For the remaining of the discussion, we will keep track of this important distinction between the properties of the individual fitted structures, and the sources as a whole. For example, we investigate the contribution of the compact radio structures to the total radio luminosity in Section~\ref{sec:differentScales}.

\subsection{Radio luminosity at 6\,GHz}\label{sec:luminosity}
\begin{figure*}
\centerline{\includegraphics[width=1\linewidth]{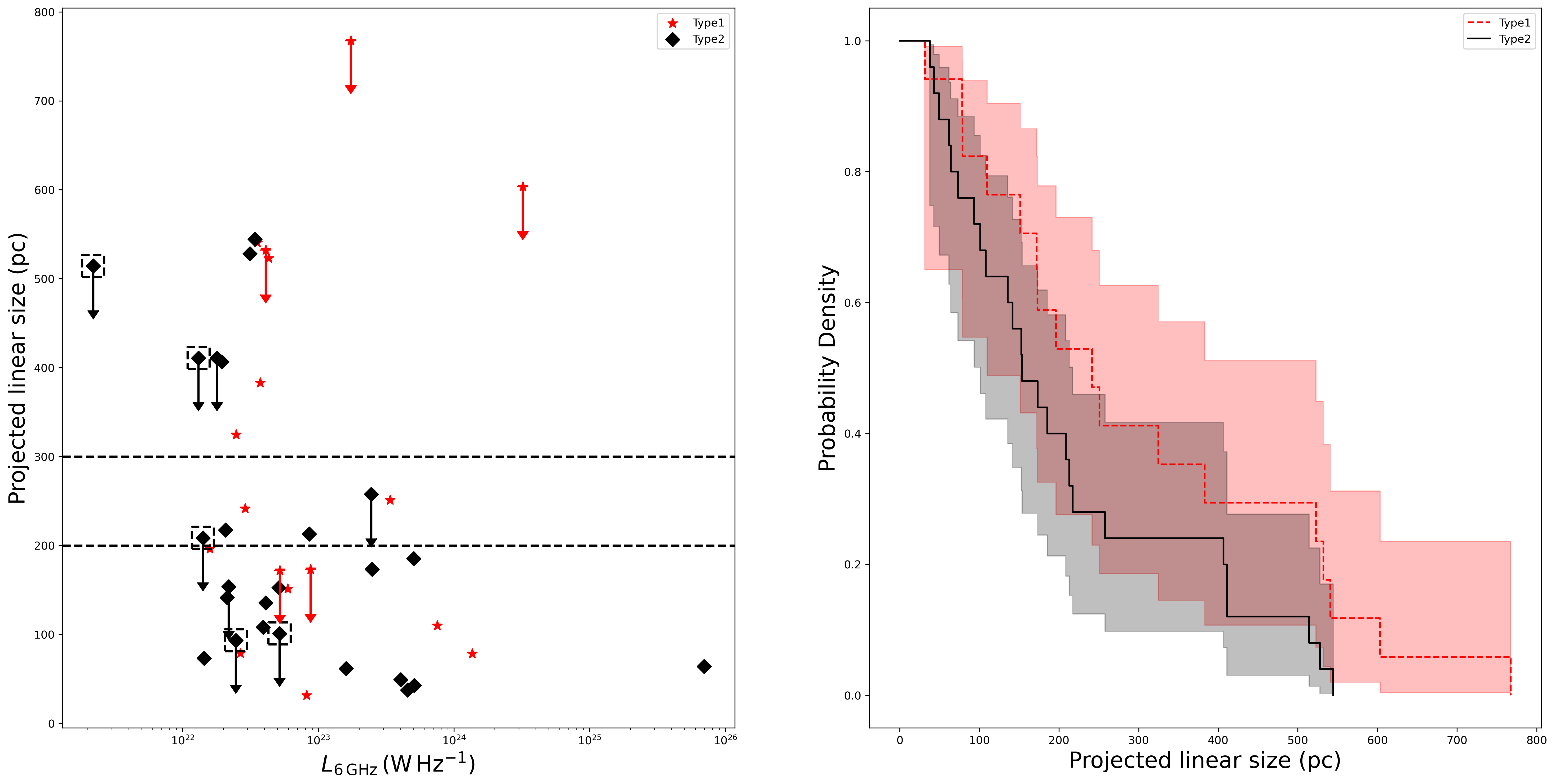}}
\caption[]{{\textit Left:} Projected linear sizes versus 6\,GHz radio luminosity for the sample in \emerlin{} (see Section~\ref{sec:luminosity}), separated into type 1 (red starts) and type 2 (black diamonds). The non-detected sources are highlighted by squares and their luminosities were derived using the lower limits on the flux density ($3 \times\,$rms). {\textit Right:} Cumulative size distributions for the projected linear sizes of the \emerlin{} radio structures for the 18 type 1 (red curve) and 24 type 2 (black curve) quasars as given by the Kaplan-Meier estimator. The shaded regions are $1\sigma$ error, and reveal no strong evidence for a difference in size distributions between the two types.}
\label{fig:LuminositySize}
\end{figure*}

Our data are insufficient to provide spectral indices of the individual structures observed in \emerlin{}. However, using the 4--8\,GHz spectral indices derived in \cite{Jarvis2021} at $\sim$0.3\,arcsec resolution VLA data (column 11 in Table~\ref{Tab:sources1}), we derived the radio luminosity $L_\mathrm{6\,GHz}$ for the structures identified (presented in column 10 of Table \ref{Tab:sources1}). In Figure~\ref{fig:LuminositySize} we plot $L_\mathrm{6\,GHz}$ as a function of the projected linear size for each target (left panel) and the size distribution of both type 1 and type 2 targets in the sample (right panel). Where more than one structure was identified, we plot the total radio luminosity of all of the structures, and used the largest projected linear size in the multi-component target. The radio luminosities range from $L_\mathrm{6\,GHz} = 9.8 \times 10^{21} \-- 7.0 \times 10^{25}\,$W\,Hz$^{-1}$, with a median of $L_\mathrm{6\,GHz} = 3.8 \times 10^{22}\,$W\,Hz$^{-1}$. 93\,per\,cent of the sources have $L_\mathrm{6\,GHz} < 1 \times 10^{24}\,$W\,Hz$^{-1}$ with only three sources (J0749+4510, J1300+0355 and J1347+1217) having $L_\mathrm{6\,GHz} > \times 10^{24}\,$W\,Hz$^{-1}$ and only 1 source, J1347+1217, showing $L_\mathrm{6\,GHz} > \times 10^{25}\,$W\,Hz$^{-1}$. This further confirms our sample is dominated by radio quiet sources (Figure~\ref{fig:QFSsample}), with moderate radio powers. We discuss the fraction of total radio luminosity located in the structures \emerlin{} maps in Section~\ref{sec:differentScales}. 

Separating our sample into either type 1 or type 2 in Figure~\ref{fig:LuminositySize} (\textit{left panel}), we find no strong evidence for a difference in the radio sizes and luminosities distribution of the radio structures identified across both types of sources. However, the type 2 quasars show a significant 0.45 Spearman rank correlation coefficient, whilst the type 1 objects show a negligible correlation of 0.02.  \cite{Thean2001} show a similar result where their type 2 sources show a significant correlation, unlike the type 1 sources. Additionally, in the right panel of Figure~\ref{fig:LuminositySize}, we constructed the cumulative distribution of projected linear sizes (including the upper limits), separated by AGN type, by using a Kaplan-Meier estimator with the y-axis showing the probability that a target is larger than a given size. Excluding the two type 1 (J0749+4510 and J1227+0419) sources showing the largest linear sizes (upper limits), there is no significant statistical differences between the two types, as is typical of most compact objects \citep[e.g.,][]{Odea1998a,Thean2001}. 




\subsection{Brightness Temperature}\label{sec:TB}
Using observables of radio flux densities and the sizes, high resolution radio imaging can be useful in disentangling star formation from AGN in radio quiet galaxies based on the brightness temperature $T_\mathrm{B}$ of a radio core \citep[e.g.][]{Condon1992,Middelberg2007}. At moderate redshifts, compact radio cores have linear sizes of $\sim$ parsec scales (corresponding to a few mas in angular size) which imply high brightness temperatures. This relates the size of the compact core to it's maximum flux density ($\theta \approx S^{0.5}_\mathrm{m}\nu^{-1}_{c}$) at GHz frequencies \citep[][]{BGW2019}. Using the observed angular sizes (intrinsic radio sizes convolved with the beam), the flux density and assuming the brightness temperature distribution of a source at a given redshift $z$ can be modelled as an elliptical Gaussian distribution \citep[e.g.][]{Condon1982,Ulvestad2005}, we derive our $T_\mathrm{B}$ following the standard equation: 
\begin{equation}\label{TBequation}
T_\mathrm{B} = 1.22 \times 10^{12}(1 + z) \left (\frac{S_\nu}{1\,\mathrm{Jy}}\right)\left(\frac{\nu}{1\,\mathrm{GHz}}\right)^{-2}\left(\frac{\theta_\mathrm{maj}\theta_\mathrm{min}}{1\,\mathrm{mas^2}}\right)^{-1}~\mathrm{K},
\end{equation}
where $\mathrm{\theta_{maj}}$ and $\mathrm{\theta_{\mathrm min}}$ are the deconvolved major and minor axes from the elliptical Gaussian model, $S_{\nu}$ is the observed flux density, and $\nu$ is the observing frequency. The derived $T_\mathrm{B}$ for \emerlin{} are shown in Table \ref{Tab:sources1}.

\begin{figure}
\centerline{\includegraphics[width=1\columnwidth]{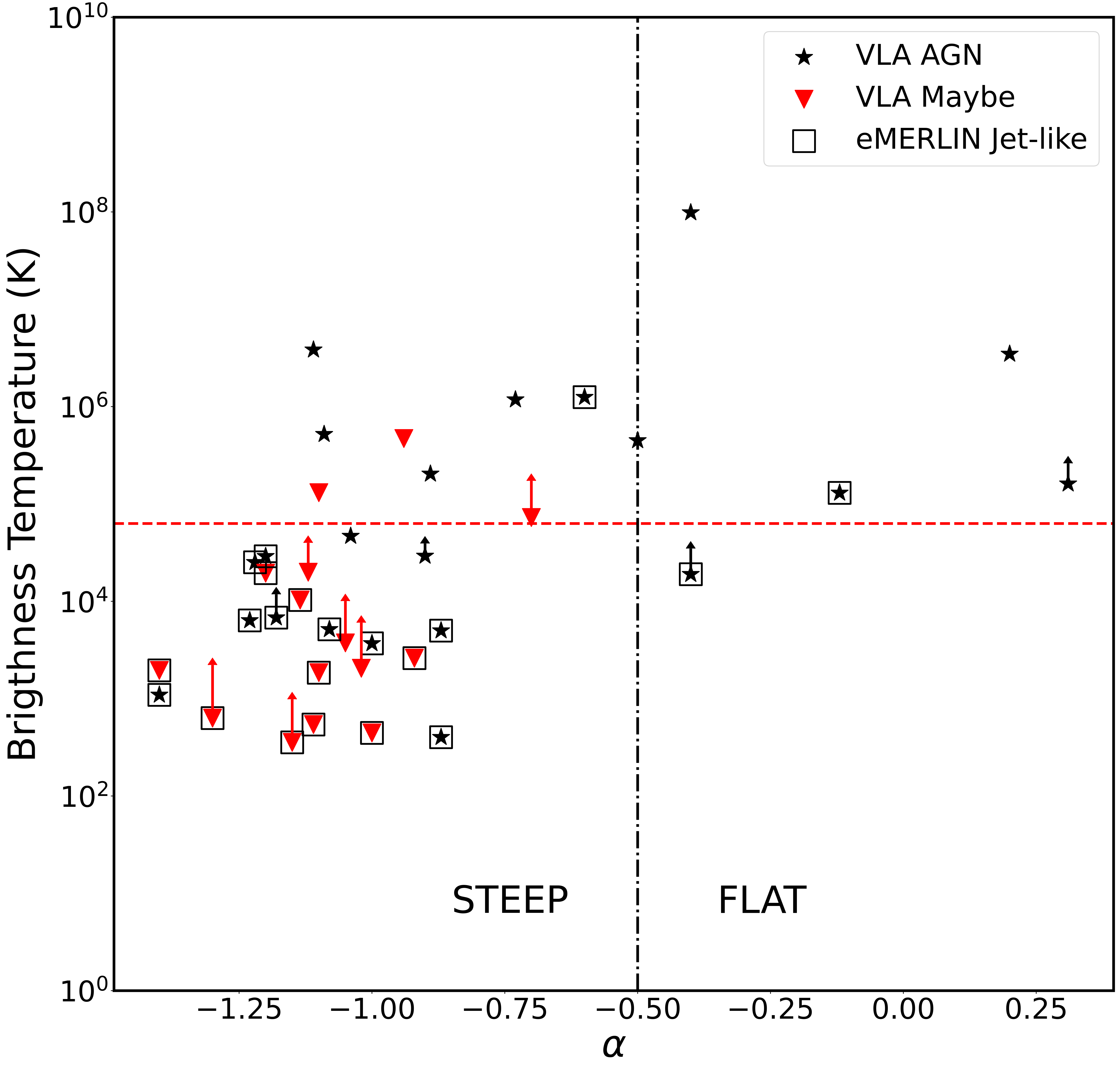}}
\caption[]{Brightness temperature ($T_\mathrm{B}$) as a function of spectral index, $\alpha$, (4--8\,GHz core spectral index from VLA) for the 37 \emerlin{} detected sources, that were either known or unknown to host radio AGN from previous VLA data (black stars and red triangles, respectively). Squares highlight sources that were classified as jet-like based on their radio morphology in \emerlin{}. For the multi-component sources, we plotted the component with the highest $T_\mathrm{B}$ value. The dashed red line highlights the lower limit of brightness temperature $T_\mathrm{B} = 10^{4.8}\,$K for selected radio AGN.}
\label{fig:TB}
\end{figure}

Figure~\ref{fig:TB} shows the $T_\mathrm{B}$ for our sample as a function of spectral index (derived from the VLA imaging; \citealt{Jarvis2021}) for the 37 targets with a detection in \emerlin{}. We note that 10 targets only have a lower limit on the brightness temperature, due to only having a upper limit on the measured sizes. Excluding the lower limits, the brightness temperature values range from $3.6 \times 10^{2} - 9.9 \times 10^{7}\,$K. \cite{Morabito2022} find $T_\mathrm{B}$ is dependent on the frequency of the observations, redshift and the spatial resolution. At a central frequency of $\sim 6.8\,$GHz and $\theta_\mathrm{res} \approx 20\,$mas, $T_\mathrm{B}>10^{4.8}\,$K will be indicative of a radio-AGN. Based on this cut-off (see dashed line in Figure~\ref{fig:TB}), 13/37 of our sources have $T_\mathrm{B}$ measurements that would be classified as radio AGN. We note that 5/13 of these sources are identified as radio-loud (see Figure~\ref{fig:QFSsample}) and therefore, all our radio-loud quasars are classified as radio-AGN based on brightness temperature. Additionally, a further 8 sources have a lower limit on $T_{B}$ that may, or may not, be consistent with the AGN cut-off. In Section~\ref{sec:radio-AGN}, we discuss the implications for identifying radio AGN across the sample and a comparison to what was previously learnt from the lower spatial resolution VLA observations of the same targets.

\section{Discussion}\label{sec:discussions}

We have presented new \emerlin{} data for the 42 $z$=0.055--0.2, type 1 and type 2, pre-dominantly `radio quiet', quasars from the Quasar Feedback Survey (sample selection in Figure~\ref{fig:QFSsample}). These 6\,GHz maps reveal a range of structures on sub-kiloparsec scales (e.g., Figure~\ref{fig:J0802+4643} and Figure~\ref{fig:sample}). In this section we discuss the implications of our key measurements of the radio morphology, radio sizes, and brightness temperatures (presented in Section~\ref{sec:analysis}). We also combine these new \emerlin{} results with previous knowledge of the kiloparsec-scale radio emission from VLA maps \citep[][]{Jarvis2019,Jarvis2021}. In Section~\ref{sec:differentScales} we discuss the radio structures on different spatial scales. In Section~\ref{sec:radio-AGN} we discuss the identification of radio AGN. In Section~\ref{sec:radio-jets} we discuss the implications of our results for AGN feedback and presented derived properties of the most clearly `jetted' systems. Finally, we explore the relationship between our sample and literature radio-identified AGN samples in Section~\ref{sec:other_populations}.

\subsection{Radio emission at sub-kiloparsec to kiloparsec scales}\label{sec:differentScales}
The combination of VLA and \emerlin{} maps enables us to trace structures from kiloparsecs to sub-kiloparsec scales. Figure~\ref{fig:J0802+4643} shows an example set of images from VLA and \emerlin{} that we have obtained for these targets. At z\,$\sim$\,0.15 (the median redshift of this sample), the 1.4\,GHz VLA maps ($\theta_{res} \sim 1\farcs0$; left panel in Figure~\ref{fig:J0802+4643}) are tracing up $\sim$10 kiloparsec scale structures, the 6\,GHz VLA maps ($\theta_{res} \sim 300\,$mas; top middle panel in Figure~\ref{fig:J0802+4643}) are sensitive to $\sim$1 kiloparsec scale structures, whilst the new 6\,GHz \emerlin{} maps ($\theta_{res} \sim 30$--100\,mas; top-right, bottom-middle and bottom-right panels in Figure~\ref{fig:J0802+4643}) are sensitive to structures ranging from 10s--100s parsec scales. 

\begin{figure}
\centerline{\includegraphics[width=1\linewidth]{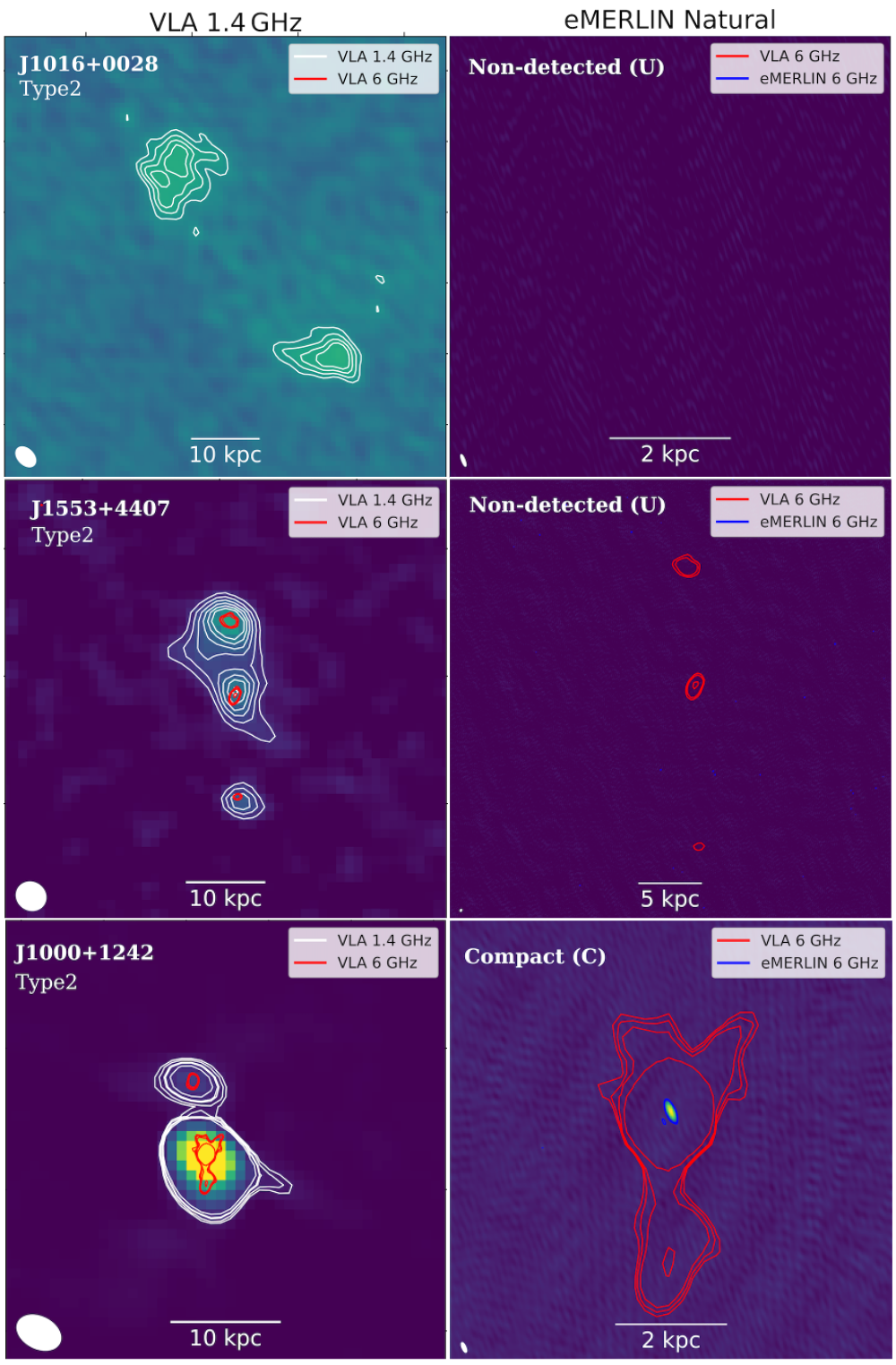}}
\caption[]{Three example targets which show AGN-like radio structures in the VLA maps (1.4\,GHz shown in the left panels and as white contours; 6\,GHz shown as red contours in all panels) but are undetected or only identified as compact in the \emerlin{} images (right panels and blue contours).}
\label{fig:LargeStructureMontage}
\end{figure}

For these targets, we find that on $\sim$10 kpc scales (traced via the 1.4\,GHz VLA maps), 80\,per\, cent of our targets reveal featureless compact structures, whereas for the $\sim 20\,$per\,cent of the sources that are partly resolved, the large scale structures are mostly resolved out at 6\,GHz ( $\sim$1 kpc scales) and are completely resolved out at sub-kpc scales in \emerlin{} (see Figure~\ref{fig:LargeStructureMontage} for examples). Therefore, it is crucial to trace the radio emission using a variety of datasets that are sensitive to different spatial scales. Otherwise, key morphological structures can be completely missed \citep[e.g.,][]{Pierce2020,Rosario2021,DAmato2022,Williams2023,Chen2024,Njeri2024}; either because the resolution is insufficient to characterise small-scale structures (affecting compact radio arrays), or the structures are diffuse and resolved away in the data (affecting extended radio arrays). Indeed, as described in this section, we see both of these regimes are extremely important for a complete characterisation of the radio emission of our quasar sample. 
Our \emerlin{} data are sensitive to sub-kiloparsec scales, with a factor of $\sim$6$\times$ and $\sim$20$\times$ higher spatial resolution than the existing 6\,GHz and 1.4\,GHz VLA maps, respectively. We see a variety of structures including weak unresolved cores, well defined hotspots/knots, sub-kiloparsec radio jets, side lobes and and irregular diffuse features, that are not identified in the lower resolution VLA images (examples in Figure~\ref{fig:J0802+4643} and Figure~\ref{fig:sample}). Such a variety of structures is consistent with previous work on quasars and low radio luminosity objects, which makes use of radio maps sensitive to sub-kiloparsec scale structures \citep[e.g.][]{Klockner2009,McCaffrey2022,Saikia2022,Baldi2023,Chen2024}.

In our \emerlin{} maps we identified 21/42 sources as having a collimated jet-like structure. Comparing these results to the VLA maps, reported in \cite{Jarvis2021}, only 7/42 sources were reported as radio-AGN based on the presence of well collimated, symmetric double or triple distinct radio peaks. \cite{Jarvis2021} was conservative in their approach of jet-like morphology classification at $\sim 1 \-- 60\,$kpc scales, and only considered the symmetric triple and double structures. However, several ambiguous sources, based on the VLA maps, are revealed to be more collimated, and clearly jet-like, at the sub-kiloparsec scales obtained in the \emerlin{} maps. These include J0759+5050 and J0958+1439, which are example sources shown in Figure~\ref{fig:sample}. Another factor that could affect the ability to resolve extended radio structures could be the orientation with respect to the line of sight. However, we note that, as shown in Section~\ref{sec:luminosity}, we find no statistically significant differences in the radio sizes between the type 1 and type 2 quasars in our sample, albeit based on a relatively small sample size. 

Our results highlight one of the key advantages of the \emerlin{} maps, enabling the identification of compact and collimated structures (along with the stronger constraints on brightness temperatures, as discussed in Section~\ref{sec:radio-AGN}). On the other hand, our new \emerlin{} maps reveal only compact cores or no detections at all, for some sources which show clearly extended structures of AGN origin (jets/lobes) in the VLA maps. We show three examples of this (J1016+0028, J1553+4407 and J1000+1242) in Figure~\ref{fig:LargeStructureMontage}, where we present the 1.4\,GHz VLA maps, the 6\,GHz VLA maps, and the new \emerlin{} natural weighted maps. The source J1016+0028 shows extended diffuse double lobe structures, separated by $\sim$40\,kpc, in the 1.4\,GHz VLA image (top left panel, Figure~\ref{fig:LargeStructureMontage}); however, the source remains undetected in both the 6\,GHz VLA and \emerlin{} data (right panel in Figure~\ref{fig:LargeStructureMontage}). The extended lobe-core-lobe radio structures in J1553+4407 as seen in both the VLA maps are completely resolved out, and remains undetected, in the \emerlin{} maps. The source J1000+1242, which is classified as having a jet based on the VLA maps, is classified as having only a compact core in \emerlin{} (bottom panels of Figure~\ref{fig:LargeStructureMontage}). This highlights that AGN radio emission can be identified on a range of different spatial scales, from sub-kiloparsecs to 10s of kiloparsecs across this sample of predominantly `radio quiet' quasars.
\begin{figure}
\centerline{\includegraphics[width=1\linewidth]{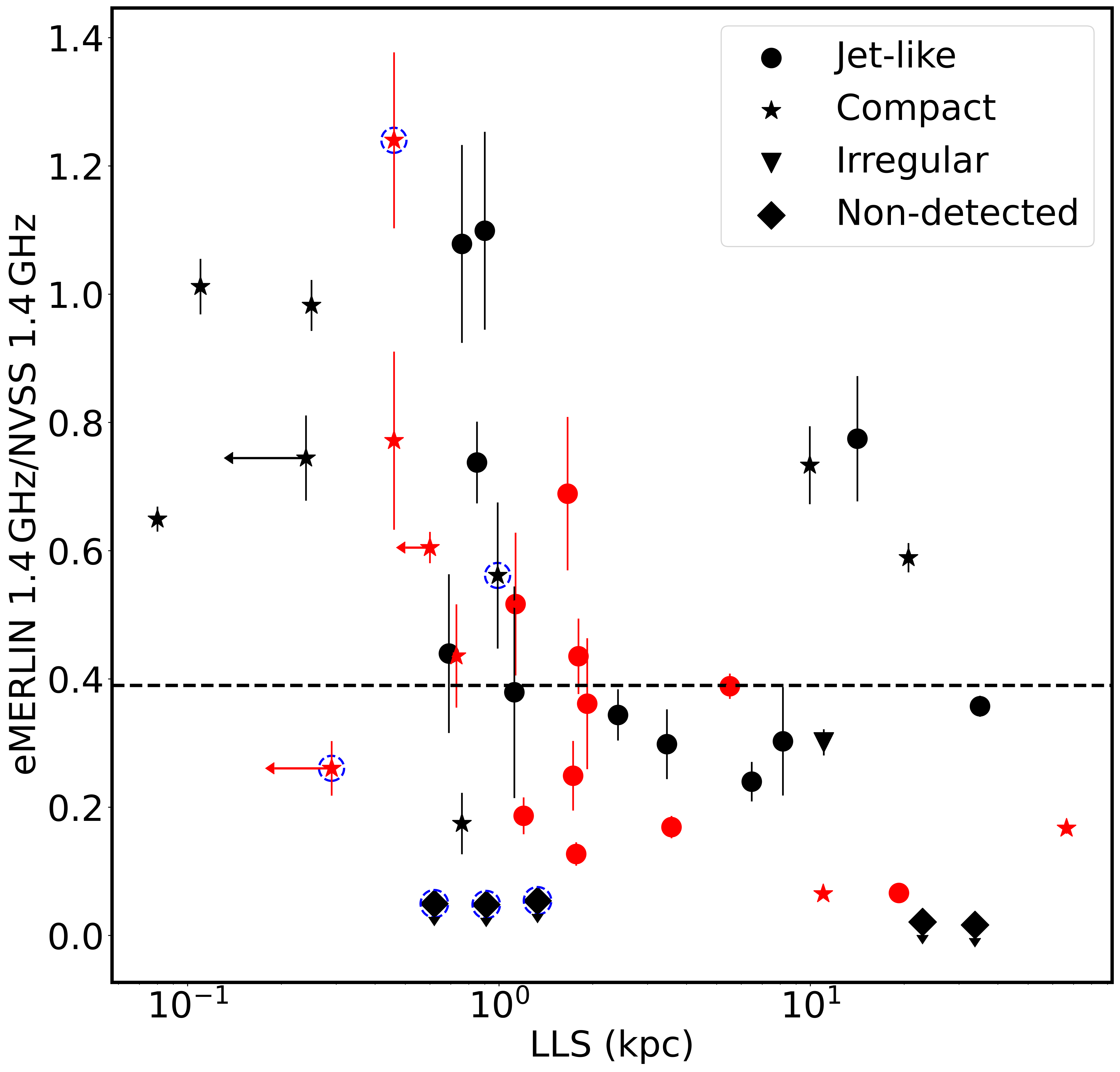}}
\caption[]{The flux density ratios of the extrapolated \emerlin{} 1.4\,GHz flux density to the NVSS 1.4\,GHz flux density as a function of the largest linear size \citep[LLS: reported in][]{Jarvis2021}. The median flux density ratio is 0.39, highlighted by the dashed line. The 6 targets highlighted in blue circles remain ambiguous (not classified as radio-AGN in this analysis or in \cite{Jarvis2021}) and are thus excluded in the calculation of the median flux density ratio.  We apply lower limits on the flux density ($3 \times \mathrm{rms}$) measurements for the 5/42 non-detected sources in the \emerlin{} data (highlighted in diamond). We also note that 2/5 non-detections are classified as radio AGN in \cite{Jarvis2021}; see Table\ref{Tab:sources1}. The legends highlight the radio morphology classification of the sources in \emerlin{}. We further split our targets into type 1 (in red) and type 2 (in black), which shows no significant variation in both the flux and largest linear size
distribution across the sample.}
\label{fig:total_flux}
\end{figure}

We can assess how much of the total radio luminosity is located in the sub-kiloparsec scale structures that we have identified in the \emerlin{} images. Under the assumption that the NVSS flux densities are a good proxy for the \textit{total} radio luminosity\footnote{We note that we are unable to account for variability between the epochs of the NVSS and \emerlin{} observations, which may particularly affect sources where the radio emission is dominated by very compact cores. Therefore, interpreting the flux density ratios for individual sources should be done with care, although we only find one such system which may have a non-physical \emerlin{}/NVSS flux density ratio $>$1.}, we extrapolated the 6\,GHz \emerlin{} flux density measurements to 1.4\,GHz to match the NVSS at 1.4\,GHz (assuming the spectral indices in Table~\ref{Tab:sources1}). The resulting total flux density ratios ($R$) of \emerlin{} to NVSS, as a function of the largest linear size (LSS) of the radio structures seen across all VLA images \citep[][]{Jarvis2021} are shown in Figure~\ref{fig:total_flux}. 

The \emerlin{}/NVSS flux density ratios have a median value of $0.39$. This implies that the sub-kiloparsec structures in \emerlin{}, only account for $\sim 40\,$per cent of the total radio emission on average. There are 7 sources which are compact based on both the \emerlin{} morphology (stars in Figure~\ref{fig:total_flux}), and clearly small in the VLA (based on LLS$\lesssim$1\,kpc) and have the corresponding, expected, high level of flux recovery in the \emerlin{} maps (R $\gtrsim$ 60\%). These sources appear to have the majority of the radio luminosity associated with very compact radio structures. Also, the most compact sources in \emerlin{} (<100\,pc) appear to have the highest radio radio luminosity (see Figure~\ref{fig:LuminositySize}, left panel). Nonetheless, only 5 sources are consistent with having $\sim$100\% of the radio emission recovered by \emerlin{}. For the sources undetected in \emerlin{}, the assumed upper limits (i.e., 3$\times$ the RMS in the maps) suggest that only $\lesssim$10\% of the emission could be located in the sub-kiloparsec scale structures. Therefore, across the sample as a whole, the majority of the high frequency (few GHz) radio emission is typically associated with structures on scales larger than the sub-kiloparsecs, to which \emerlin{} is sensitive. This result may be consistent with the idea that the highest accretion rate radio quiet sources (like this sample), typically become less core dominated (\citealt{Alhosani2022}); however, a large systematic survey across a wider range of accretion rate systems is required.
 
Overall, we see a huge range in morphological structures in the quasar sample, including sources which are dominated by compact cores, sources with $\lesssim$1\,kiloparsec scale jet-like emission and sources with $\gtrsim$10\,kiloparsec scale radio lobes. Importantly, Figure~\ref{fig:total_flux} highlights, that even for sources which appear compact in \emerlin{}, the majority the radio emission can be located in extended structures. This adds to the growing evidence that radio emission associated with the radio-quiet quasar population is a complex mix of structures/processes, spanning a range of galactic and sub-galactic scales \cite[e.g.,][]{Jarvis2019,Panessa2019,Smith2020a,McCaffrey2022,Silpa2022,Silpa2023}. In the following section we investigate how many of the sources have radio emission that can be attributed to AGN-related processes. 


\subsection{Identification of Radio AGN}\label{sec:radio-AGN}
As radio quiet quasars do not produce powerful relativistic jets, the origin of the relatively weak radio emission in quasars still remains largely unknown \citep[e.g.][]{HeckmanBest2014,Kellerman2016,Maini2016,Panessa2019}. This weak radio emission might arise from the AGN, star formation processes or both of these. A direct detection of collimated jet-like emission, or a bright and central compact radio core with high brightness temperatures (Section~\ref{sec:TB}) can serve as an AGN identifier. We used radio morphology classification and brightness temperature measurements to investigate the origin of radio emission in these quasars. 

Figure~\ref{fig:TB} shows the brightness temperatures for our \emerlin{} detected targets as a function of the spectral index measured in \cite{Jarvis2021}. At least 13, and a maximum of 21 (accounting for lower limits), of these have brightness temperatures above the limit of $T_\mathrm{B} = 10^{4.8}\,$K, which is the limit assumed to be attributable to AGN emission (Section~\ref{sec:TB}). Three of the confirmed sources (highlighted as three red triangles above the $T_\mathrm{B} = 10^{4.8}\,$K cutoff in Figure~\ref{fig:TB}) were not previously identified as AGN based on the VLA analysis presented in \cite{Jarvis2021}. This is because the VLA constraints on the brightness temperatures were below the required threshold. This highlights that \emerlin{}'s superior resolution helps identify some AGN cores. Nonetheless, this is only three sources, despite the order-of-magnitude increase in spatial resolution between \emerlin{} and VLA. Indeed, our work indicates a low level of unambiguous AGN cores ($\sim$31--50\%) across the sample, with a large fraction of the radio emission located within larger-scale structures (Section~\ref{sec:differentScales}). Further work using very high resolution interferometry, and supplemented by spectral index constraints, will help determine the prevalence of true AGN radio cores across the radio quiet quasar population as a whole \citep[e.g.,][]{Chen2023,Chen2024b}.

For the sources without a high brightness temperature AGN core, we also rely on morphology to identify radio AGN. The sources that show jet-like structures still show diversity, as shown in Figure~\ref{fig:sample}. Indeed, many low-luminosity radio galaxies are characterized by diverse morphologies \citep[e.g.][]{Leipski2006,Luo2010,Odea2021}. Multi-component sources with symmetric double-lobed morphology and well collimated radio knots are indicative of AGN jets, or possibly shocks from outflows, indicating the presence of a radio AGN \citep[e.g.][]{Leipski2006,Middelberg2007,Odea2021,Fischer2023}. Therefore, the 21/42 sources, which were classified as jet-like based on their morphology, are assumed to host a radio AGN. These are highlighted with squares in Figure~\ref{fig:TB}. 

Combining both morphology and brightness temperatures, we classified 32/42 sources as hosting a radio-AGN based on the \emerlin{} data alone. There are 4 additional sources (J1010+1413, J1016+0028, J1100+0846 and J1553+4407) which were identified as hosting a radio AGN based upon the VLA data in \cite{Jarvis2021}, but not in the \emerlin{} data. Of these, two (J1016+0028 and J1553+4407) remain undetected in \emerlin{} but their radio AGN clasification is primarily based on the larger scale radio morphology only seen in VLA maps (see Figure~\ref{fig:LargeStructureMontage}). The other two sources (J1010+1413 and J1100+0846) were identified as radio-AGN based on a `radio excess' above the far-infrared radio correlation and further evidence of radio AGN is presented in \cite{Jarvis2019,Jarvis2020}. Combining the new information from \emerlin{}, and that obtained with the VLA images, we identify radio emission associated with an AGN in 36/42 ($\sim$86\,per\,cent) of the sources in our Quasar Feedback Survey. This is a significant increase on the $\sim 57\,$ per\,cent of the sources identified using only the VLA data in \cite{Jarvis2021}. Such a large fraction of radio AGN, not easily identified using traditional methods (typically with spatially-unresolved datasets), could have important implications for obtaining a complete census of radio AGN activity, and for de-contaminating AGN in radio studies of star-forming galaxy populations \citep[][]{Jarvis2021,Hansen2024}.

 
Although a radio AGN is identified in $\sim$86\,per\,cent of the quasars in our sample, it does not mean it is the \textit{dominant} radio emission mechanism in all the targets. With \emerlin{} typically recovering only $\sim 40\,$per\,cent of the total flux density on average (Section~\ref{sec:differentScales}), this implies that more than $60\,$per\,cent of the total radio emission in the sample can be attributed to extended radio emission. In the absence of other information, the radio emission that is completely resolved away could be associated with star formation or diffuse AGN outflow-driven shocks. Indeed, a significant fraction of diffuse, extended radio structures has been previously noted in quasar samples \citep[][]{Kimball2011b,Condon2013,McCaffrey2022}. In the following sections, we discuss the radio emission from these quasars for understanding AGN feedback processes.



\subsection{Implications for feedback and radio jets at sub-kiloparsec scales}\label{sec:radio-jets}
Typically at the moderate luminosities, which represents our sample (Figure~\ref{fig:QFSsample}), i.e., intermediate between `radio loud' and very `radio quiet', star formation cannot be the dominant mechanism contribution to the radio emission \citep[e.g.,][]{Zakamska2016,White2017,Jarvis2019,Silpa2023,Liao2024}. Therefore, other AGN-related processes such as compact, low power radio jets and/or shocks due to AGN-driven outflows are expected to dominate \citep[e.g.,][]{Orienti2010,Nims2015, Zakamska2014,Kharb2021,Yamada2024,Harrison2024}. Distinguishing these different processes with radio morphology alone can be very challenging, and additional information such as polarisation might be crucial \citep[e.g.,][]{Silpa2022,Fischer2023,Meenakshi2024}. Nonetheless, the indication is that the radio emission is tracing an interaction between the AGN and the host galaxy ISM. 

Indeed, nine targets from this study have already been observed to show a spatial alignment between the radio emission and multi-phase outflows on kiloparsec scales, as traced in [O~{\sc iii}]
and CO, indicating that the radio emission is tracing sites of ISM interactions \citep[][]{Jarvis2019,Girdhar2022,Girdhar2024}. Furthermore, there is now a lot of evidence in the literature (both from spatially-resolved observations and statistical samples) that radio emission is a good tracer of the energy deposited by AGN into the ISM of their host galaxies across the quasar populations \citep[e.g.,][]{Zakamska2014,Harrison2015,Odea2021,CalistroRivera2024,Petley2024,Harrison2024}. Intrinsically small radio sources (sub-galactic scales), which dominates this sample, may be at the early stages of evolution, and consequently the most likely to observe such ISM interactions \citep[e.g.,][]{Odea1998a,Bicknell2018,Santoro2020,Odea2021,Morganti2023,Kukreti2024,Marques2024}. The $100$\,pc scale structures identified in \emerlin{} in this study, would therefore advocate for comparable resolution observations of the multi-phase gas properties of this sample. Indeed, local studies imply that AGN impact could commonly be happening on these $\sim$100\,pc scales \citep[e.g.,][]{Venturi2021,Singha2023,Holden2024,Haidar2024}.

\subsubsection{Jet properties}

We now continue with the assumption that the majority of the sources in our sample contain low power jets (but see discussion above). The causes for the less-powerful radio jets to be compact, and less-collimated, has been attributed to several factors. This includes a lack of sufficient energy to dissipate efficiently and thus remain entrapped within the host galaxy; the jets are frustrated and their propagation interrupted by the dense interstellar medium \citep{Van_Breugel1984,Bicknell2018,Mukherjee2018,McCaffrey2022,Kharb2023}; or the jets being sub-relativistic and thus unable to grow due to low jet bulk velocity, or more easily become disrupted \citep{Perucho2014,Giovannini2023,Dutta2024}. The exact mechanism driving these moderate jets in these radio-quiet quasars remain uncertain with magnetic fields, black hole spin and accretion disk winds thought to play a key role \citep[e.g.][]{Reeves2009,Shen2014,Panessa2019,Rusinek2020}.

These low power compact jets, still could have sufficient energy to disrupt the ISM and affect the host galaxy evolution, depending on their relative inclination angle with respect to the host galaxies \citep[e.g.,][]{Mukherjee2018,Mandal2021,Webster2021,Tanner2022,Marques2024}. Indeed, if low-powered jets remain trapped in the ISM for a longer time, they interact with the ISM over a large volume, evacuating cavities upto $r\sim 1\,$kpc and only leaving the densest cores (\citealt{Mukherjee2016}). Therefore, studying these low-powered jets ($P_\mathrm{jet} \lesssim 10^{44}$\,erg\,s$^{-1}$) are crucial in our understanding of the feedback mechanisms at galactic scales \citep[e.g.][]{Harrison2015,Kharb2023}.

\begin{figure*}
\centerline{\includegraphics[width=1\linewidth]{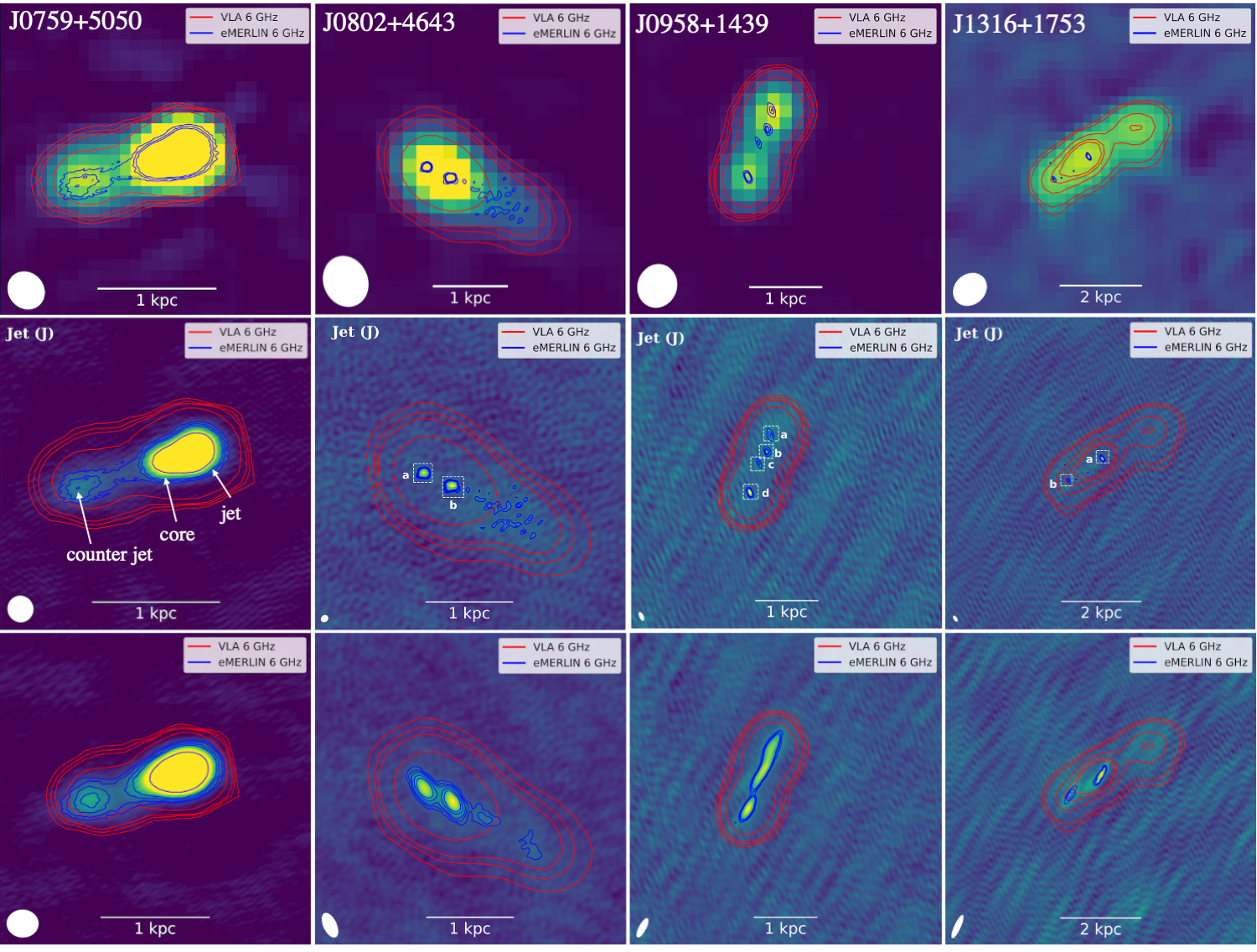}}
\caption[]{The four targets with the most clearly defined jetted emission, which enable common jet properties to be derived. The top row shows the VLA 6\,GHz images, where the blue contours are from the \emerlin{} natural-weighted images. The VLA 6\,GHz images are represented by red contours in all panels. The middle row shows the natural-weighted \emerlin{} images and the bottom row shows the \emerlin{} tapered images.  The components used for calculations are labelled in the middle row (see Section~\ref{sec:radio-jets}).}
\label{fig:JetMontage}
\end{figure*}

With all this in mind, we estimate jet properties (speeds, inclination angles, and powers) using our \emerlin{} maps, for the systems where this is possible. The disrupted nature of low power jets means that standard approaches to calculate jet powers and inclination angles can be challenging, due to a lack of clearly defined structures. Nonetheless, aligned multi-components in radio quiet quasars are also thought to be termination points of radio jets \citep[e.g.][]{Leipski2006}, and we identify the four targets with the strongest evidence for such structures, which allow us to perform typical types of jet measurements done in the literature. These are the four sources shown in Figure~\ref{fig:JetMontage}. 

We estimate the jet kinetic power for all the four targets using the jet radiative efficiency empirical formula of \cite{Merloni2007}:
 \begin{equation}
     \mathrm{log}\,P_\mathrm{jet} = (0.81+0.11) \times \mathrm{log}\,L_\mathrm{6\,GHz} + 11.9^{+4.1}_{-4.4}
 \end{equation} 
 
We derive jet-to-counterjet surface brightness ratios, $R_J$, for these sources, assuming that the asymmetries in surface brightness are due to Doppler boosting and dimming effects in approaching and receding jets, respectively. We also calculate the jet inclination angles using the K-corrected radio core prominence parameter, $R_\mathrm{C}$, following equation (C2) in \citet{Urry1995}; it is assumed that the core has a spectral index of 0 and the extended emission of $-1$.
$$R_\mathrm{C} = (S_\mathrm{core} / S_\mathrm{ext})(1+z)^{-1} = (S_\mathrm{core} / (S_\mathrm{tot} - S_\mathrm{core}))(1+z)^{-1}.$$ 
For the estimation of the K-corrected $R_\mathrm{C}$, we have used the core flux density from the \emerlin{} images while the total flux density was estimated using the VLA images. The {\tt AIPS} verbs {\tt TVMAXF} and {\tt TVSTAT} were used to derive the core and total flux densities, respectively. Following the work of \citet{Kharb2004}, we have attempted to use the $R_\mathrm{C}$ parameters to determine jet inclinations in the four jetted sources. \citet{Kharb2004} had estimated $R_\mathrm{C}$ values for a large sample of both plane-of-sky radio galaxies as well as their pole-counterparts, i.e., the blazars, and found that the log\,$R_\mathrm{C}$ values varied roughly from $-4$ to $+4$ for jet inclinations respectively going from $90^{\circ}$ to $0^{\circ}$. Furthermore, in the absence of multi-epoch observations to calculate proper motions, using the $R_J$ values, we have derived jet speeds using equation (A10) in \citet{Urry1995}; that is, $\mathrm{\beta cos \theta = \frac{{R_J}^{1/p} - 1}{{R_J}^{1/p} + 1}}$. The jet structural parameter, $p$ is assumed to be = 3 for a `continuous' jet. We apply these calculations to the four most applicable systems, as follows:

\begin{enumerate}
\item J0759+5050 shows a two-sided jet structure. We derived a K-corrected log\,$R_\mathrm{C}$ value of $\sim-1.2$ ($S_\mathrm{core}=866~\mu Jy, S_\mathrm{total}=12.95~$mJy). We note that this source in particular did not show a clear compact `core' in its \emerlin{} image. We used the {\tt AIPS} verb {\tt CURV} to read off the flux density of the brightest pixel. A jet inclination of $60\degr$ is implied. Using the emission from a `hotspot' in the brighter jet towards the north-west and a similar `hotspot' in the counterjet towards the south-east, we derive an $R_J = 15.5$. For an inclination angle of 60$\degr$, a jet speed of 0.85c is needed to explain the $R_J$ value. We estimate a jet power of $\sim$10$^{45.4}$\,erg\,s$^{-1}$. 

\item J0802+4643, shows two bright and compact components and an aligned plume of diffuse emission extended to the south west. In this case, component `b' is assumed be the core and component `a' a hotspot in the approaching jet. We derive a K-corrected log\,$R_\mathrm{C}$ value of $\sim-0.8$ ($S_\mathrm{core}=467~\mu Jy, S_\mathrm{total}=3.13$~mJy) and a jet inclination of $55\degr$. Using the counterjet surface brightness at a distance roughly equidistant from the core, a jet-to-counterjet surface brightness ratio can be derived; this turns out to be $R_J = 3.2$. For the jet inclination angle of $55\degr$, this would imply a jet speed of 0.33c. We estimate a jet power of $\sim$10$^{43.8}$\,erg\,s$^{-1}$. 
    
\item J0958+1439, shows a double symmetric radio structure in VLA images. At the sub-kiloparsec scales of \emerlin{}, the extended emission is  resolved out revealing 4 faint but distinct components. We note that the tapered image shows a well collimated radio jet connecting the northern components (a,b,c) to the southern component (d). Component `d' is assumed to be associated with the approaching jet and `a' with the counterjet. We derive a K-corrected log\,$R_\mathrm{C}$ value of $\sim-1.2$ ($S_\mathrm{core}=157~\mu Jy, S_\mathrm{total}=2.32$~mJy) and a jet inclination of $60\degr$. We also derive an $R_J = 1.7$, which would suggest a jet speed of 0.18c for the jet inclination angle of $60\degr$. We estimate a jet power of $\sim$10$^{43.5}$. 

\item J1316+1753 shows two lobes separated by $\sim 2\,$kpc in the VLA images, whereas these side-lobes are resolved out at sub-kiloparsec to reveal two radio knots. Component `a' is assumed to be the core, knot `b' is jet and counterjet is not clearly detected. Therefore, we can derive a lower limit to the jet-to-counterjet surface brightness ratio of $R_J \ge2.4$. We derive a K-corrected log\,$R_\mathrm{C}$ value of $\sim-0.8$ ($S_\mathrm{core}=337~\mu Jy, S_\mathrm{total}=2.17$~mJy) and a jet inclination of $55\degr$. For this inclination angle, the lower limit to the jet speed would be 0.25c. We note that the jet inclination derived here differs from the $45\degr$ derived for J1316+1753 earlier by \citet{Girdhar2022}. This is because \citet{Girdhar2022} had used the 5.2~GHz VLA core flux density for estimating the $R_\mathrm{C}$ without K-correction; the \emerlin{} core flux density used here is less contaminated by jet emission close to the core. We estimate a jet power of $\sim$10$^{43.6}$\,erg\,s$^{-1}$. 
\end{enumerate}

Three of these four targets have moderate inferred jet speeds of $\sim$(0.2--0.3)\,c, which have indeed been derived or observed to be common in other radio-quiet quasars in the literature \citep[e.g.,][]{Sbarrato2021,Wang2021,Wang2023}.  While their radio-loud counterparts often produce highly relativistic speeds, the jets in radio-quiet AGN are generally slower and high resolution radio imaging have revealed jet components moving at sub-relativistic speeds in these objects, and could even be regarded as outflows or `winds' \citep[][]{Stocke1992,Wang2021,Fischer2023}. Future work comparing such derived jet properties to measurements of the host galaxy ISM, will help assess the  efficiency of jet-ISM interactions \citep[e.g., following][]{Girdhar2022}.


\subsection{Comparison to other radio AGN populations}\label{sec:other_populations} 
We have identified a large fraction of radio AGN in our sample of optically selected type 1 and type 2 quasars. Our sources are predominantly compact on $\lesssim$1\,kpc scales. Therefore, we discuss the relationship between our quasar sample to other similar compact radio-emitting AGN such as Seyferts, compact steep-spectrum sources (CSS), gigahertz-peaked spectrum (GPS) sources and FR0s \citep[][]{An_Baan2012,Odea2021,Baldi2023}. At a few $\sim$GHz frequencies, the \emerlin{} has sufficient resolution to probe Seyferts (and other low luminosity radio AGN) due to their faint nature (at a few millijanskys) and $<1\,$kpc in size, by minimising the contribution of the host galaxy and resolving out sub-kiloparsec to parsec-scale radio cores and knots as seen in this work and other similar probes \citep[e.g.][]{Thean2000,Baldi2021b,Baldi2021c}. 

For example, FR0s sources have similar properties to FRIs but are a factor of $\sim 30$ more core-dominated than FRIs and remain featureless on kiloparsec scales \citep[][]{Baldi2016,Hardcastle2020,Baldi2023}. Indeed, comparing the radio morphology of our sample to the radio morphology of FR0s sources in the literature, at several kiloparsec scales, most of our sample remains compact and unresolved as is the case with FR0s. However, at a few kiloparsec (1\--3\,kpc) scales, a third of the sample shows one-sided or two-sided core-jets with $\sim 80\,$per\,cent of the sources showing steep spectrum ($<-0.7$) in VLA \citep[][]{Jarvis2021}. A similar result is reported in \cite{Baldi2015}, \cite{Baldi2019} for FR0s, where only a third of their sample exhibits double or one-sided radio jets at an angular resolution of about $0\farcs3$. Furthermore, at sub-kiloparsec scales, FR0s show a range of diverse radio properties and radio morphologies from steep to flat spectral indices and multi-components with luminosities $10^{22}\-- 10^{24}\,$W\,Hz$^{-1}$ \citep[][]{Baldi2016}. At these sub-kiloparsec scales probed by \emerlin{}, our sample mainly shows compact radio jets (one-sided or multi-component radio knots) at $\sim <0.5\,$kpc or unresolved compact cores with radio luminosities similar to the ones reported in \cite{Baldi2016}. Two-thirds of our sample appear to be slightly core dominated at $\geq 1/3$ of the total emission (see Figure~\ref{fig:total_flux}). \cite{Baldi2021c} combine both the VLA and \emerlin{} data of similar objects to reveal sub-kiloparsec jets and conclude that FR0s harbour sub-kiloparsec scale radio jets whose low surface brightness make it difficult to detect and isolate in typical datasets.

Our targets have projected linear size of $<500\,$ pc in \emerlin{}, which covers the GPS size region \citep[see reviews by][]{Odea1998a,Odea2021}. Furthermore, following the power versus linear size diagram of \cite{Hardcastle2020} and \cite{Jarvis2019}, our quasars lie on the lower end of the radio power for the GPS and CSS objects. The two multi-component sources, J0802+4643 and J1317+1753 (see Figure~\ref{fig:JetMontage}), show compact symmetric radio structure with well-defined and confined side lobes. The distances between the two lobes is $365\,$pc in J0802+4643 (also see Figure~\ref{fig:J0802+4643}) and $1\,$kpc in J1317+1753, with spectral indices of $\alpha \sim -1.4$ and $-1.2$, respectively (from \citealt{Jarvis2021}). Based on the double-lobe radio morphology, the projected linear size of $<1\,$kpc and the steep spectra, these two sources would qualify as GPS/CSS objects. The source J0802+4643 is a likely GPS ($ D \sim 364\,$pc) source with a frustrated jet within the ISM as indicated by the plume-like diffuse radio emission located at a considerable distance away from the two hotspots \cite[e.g.][]{An_Baan2012,Odea2021}. On the other hand, the source J1317+1753 is a likely CSS object ($D\sim 1\,$kpc) representing the young revolutionary phase of a galaxy that might later evolve into an FR-II source as reported in \cite{An_Baan2012,McCaffrey2022}. Whilst probed at $\sim6\,$kpc, \cite{McCaffrey2022} shows a similar (and first) FRII-like radio quiet quasar with $L_\mathrm{6\,GHz}\approx 10^{22.9}\,$W\,Hz$^{-1}$ at kiloparsec scales. We also note that side lobes remained unresolved at $\sim$kpc scales and due to lack of ancillary data, the source was not classified as radio-AGN in \cite{Jarvis2021}. Additionally, the most luminous object (J1347+1217) in our sample at $L_\mathrm{6\,GHz} \sim 7.0 \times 10^{25}\,$W\,Hz$^{-1}$, is a confirmed GPS source and is reported as a rejuvenated radio object \citep[][]{Morganti2013,Odea2021}. 

The extended radio structures of radio-quiet quasars at sub-arcsecond resolution are by no means different from that of Seyferts as reported in \cite{Leipski2006}. Among the luminosities studied here, the morphological features and radio sizes are found to be similar in both types of objects and there is no significant differences between the radio luminosities of the type 1 and type 2 sources as shown in Figure~\ref{fig:LuminositySize}. Although, type 2 sources show a significant correlation between radio power and radio sizes, with these targets showing evidence of collimated radio structures that are also reported in \cite{Thean2001}. We note that all our multi-component sources (in Figure ~\ref{fig:JetMontage}) showing well collimated radio structures are of type 2 and radio-quiet quasars. Whilst type 1 sources show no significant correlation, we also note that a large fraction of our type 1 sources are classified as 'compact' based on their radio morphology. There is no statistical difference in the size distribution of both type 1 and type 2 sources, indicating that the central engines for each type are similar and both equally contribute to the extended emission \citep[][]{Thean2000,Thean2001,Pierce2020}. This supports the argument in the literature that radio-quiet quasars are scaled-up versions of Seyfert galaxies \citep[e.g.,][]{Kukula1998,Chiaraluce2019}. Furthermore, \cite{Baldi2021b,Baldi2021c} show that the compact and edge-brightened radio jet morphology observed in Seyferts can be attributed to the low-powered nature of the slow jets and disk/corona winds in disk/spiral galaxies that host lower spinning black holes, which hinder launching of faster jets. Therefore, the relatively small amount of radiative energy ($P_\mathrm{jet} \approx \times$10$^{43}$\,erg\,s$^{-1}$) emerging at radio wavelengths in these radio quiet quasars, may simply be due to the presence of less powerful radio jets \citep[][]{Ulvestad2005}. 

As is typical across the quasar population, $\sim 90\,$per\,cent of our sample is radio-quiet based on traditional definitions. However, these prevalent radio-AGN appear to be a natural extension of `high-luminosity' for Seyferts and `low-luminosity' for CSS/GPS objects \citep[also see Fig.2 in ][]{Hardcastle2020}. Similarly, \cite{Chilufya2024} find that their sample of low-luminosity radio-loud AGN share the same region in the radio luminosity versus size diagram with radio-quiet quasars and FR0s, and show similar properties in spectra to GPS and CSS objects. As the literature shows, based on samples selected from a variety of different criteria, it is increasingly difficult to separate these radio populations into distinct classes based on their radio morphology and luminosity \citep[e.g.][]{Mingo2019,Harwood2020,Capetti2020,Pierce2020,Kumari2021,Gurkan2021,Mingo2022,Stroe2022,Chilufya2024}. Perhaps a more general question is if these radio quiet quasars (like those studied here) show similar radio properties to other low-luminosity compact radio objects and to their higher luminosity counterparts. Future studies could focus on spectral energy distributions and polarisation measurements to further probe the differences, if any, across these low-luminosity compact radio AGN populations \cite[e.g.,][]{Jeyakumar2016,Silpa2022,Meenakshi2024}.

%




\section{Conclusion}\label{sec:conclusion}
We present a sample of 42 low redshift $z<0.2$ type 1 and type 2 quasars observed with \emerlin{} at 6\,GHz. These targets have moderate radio luminosities, and the sample is dominated by `radio quiet' quasars, based on the [O~{\sc iii}] to radio luminosity ratio (Figure~\ref{fig:QFSsample}). The new \emerlin{} data enabled us to measure radio structures on $\sim$\,10s-100s\,pc scales, which is an order of magnitude smaller than the structures measured in the existing VLA images of these targets (example in Figure~\ref{fig:J0802+4643}). Our main conclusions are:

\begin{itemize}
    \item The targets show a wide range of morphology on sub-kiloparsec scales, with compact cores, knots, extended collimated structures and more diffuse structures. Based on visual inspection only, 15/42 (35\%) targets are classified as compact sources as they show no extended radio features, 21/42 (50\%) targets are classified as jet-like since they show either one-sided jets or well collimated multiple radio components and 1/42 source has an irregular morphology. 5/42 targets were not detected in the maps and 5/37 of the detected sources show multiple distinct (2--4) radio peaks and are thus referred to as multi-component sources (Figure~\ref{fig:sample}). 
    \item We measured the properties of the structures identified in the \emerlin{} maps. The projected linear sizes range from, $\sim$30–540\,pc, with a median size of $\sim$147\,pc, excluding the 10 components which remained unresolved on these scales (Figure~\ref{fig:flux}). These structures typically have modest luminosities, with $L_{\rm 6GHz}\sim10^{22}$--10$^{24}$\,W\,Hz$^{-1}$ (with only 2 higher), and we found no evidence of significant differences in these properties between the type 1 and type 2 quasars (Figure~\ref{fig:LuminositySize}). 
    \item Although there is a large variety across the sample, we found that a significant fraction of the total radio luminosity in this sample is typically not located in the compact, sub-kiloparsec structures traced by \emerlin{}. Indeed, most targets have moderate brightness temperatures, with only 13--21 targets showing clear evidence for AGN-like values (i.e., $T_{B}>10^{4.8}$\,K; Figure~\ref{fig:TB}). This corresponds to 31--50\%, with the range due to some targets only having an upper limits on their sizes. Furthermore, under the assumption that NVSS traces the total radio luminosity, we find that, on average, $\sim60$\% of the radio luminosity is resolved away on sub-kiloparsec scales traced by \emerlin{} (Figure~\ref{fig:total_flux}). Previous VLA maps of the same targets show that this resolved-away emission can be sometimes located in diffuse, loosely collimated lobes, likely associated with AGN jets or outflows (Figure~\ref{fig:LargeStructureMontage}). 
    \item By combining diagnostics of brightness temperature and morphology, we identified radio AGN activity in 32/42 (76\%) of the sample (Figure~\ref{fig:TB}). This is an increase on $\sim 57\,$per\,cent, based only on the VLA maps, thanks to the increased constraints on morphology and brightness temperatures at the higher resolution of \emerlin{}. Nonetheless, 4 additional targets are only identified as AGN based on the VLA maps (not \emerlin{}) due to large-scale structures being resolved away (Figure~\ref{fig:LargeStructureMontage}), bringing the total number of radio AGN being identified in the sample to be 36/42 (86\%).  
    \item For the four targets which showed sufficiently defined jet structures (see Figure~\ref{fig:JetMontage}), we estimated the jet properties. For three sources, the jet speeds and powers are estimated to be modest, with values of $\sim$0.2--0.3\,c and $P_\mathrm{jet} \approx$ (3--6)$\times$10$^{43}$\,erg\,s$^{-1}$, respectively. If jets are driving the radio emission across the majority of the sample, this implies modest jet speeds and modest powers are common. The lack of defined structures in the majority of the sample (cores, jets, counter-jets) could be partly related to disruption/dissipation by ISM interactions, or a larger contribution to the radio emission from outflows launched by wide-angle winds.
\end{itemize}

Our results highlight the importance of using a wide range of radio images (tracing sub-kiloparsec scales through to 10s of kiloparsec scales), to obtain a full characterisation of the radio morphologies in a representative sample of quasars. Through these data, we find a high prevalence of radio emission associated with an AGN ($\sim$90\%) for a sample which is traditionally characterised as `radio quiet'. The sample shares many radio properties with other AGN populations, including Seyferts AGN, FR0s, and GPS/CSS sources; highlighting limited evidence for these being distinct populations. 

This work adds to the growing evidence, by studying low-redshift systems, the importance of using sensitive radio imaging, down to sub-kiloparsec scales, to study AGN feedback mechanisms on galactic scales, even in low--moderate radio luminosity systems. We advocate for similar work in objects at cosmic noon, where the bulk of black hole growth occurs. This will be important to establish the driving mechanisms of multi-phase outflows that are now commonly identified at cosmic noon, but the bolometric AGN output is often assumed to be the most important driving factor. Indeed, combining JWST data with radio imaging already shows a radio-outflow connection in `radio quiet quasar' (following our definition) at $z\sim$1.5 \citep[][]{Cresci2023}. Ongoing work to provide sensitive and high resolution radio imaging of large samples \citep[e.g.,][]{Muxlow2020,Morabito2022,Njeri2024}, and future work with new facilities \cite[e.g., the Next Generation VLA and the Square Kilometre Array;][]{Nyland2018}, will be crucial to characterise the sub-galactic scale radio emission and compare to spatially-resolved multi-phase gas observations. 



\section*{Acknowledgements}
We thank the anonymous referee for their constructive report. We thank the \emerlin{} support staff for assistance with these observations. \emerlin{} is a National Facility operated by the University of Manchester at Jodrell Bank Observatory on behalf of STFC, part of UK Research and Innovation. AN and CMH acknowledge funding from an United Kingdom Research and Innovation grant (code: MR/V022830/1). SM acknowledges funding from The Royal Society through a Research Fellows Enhancement Award Grant (RF/ERE/221053). PK acknowledges the support of the Department of Atomic Energy, Government of India, under the project 12-R\&D-TFR-5.02-0700. SS acknowledges financial support from Millenium Nucleus NCN23\_002 (TITANs) and Comit\'{e} Mixto ESO-Chile.

\section*{Data Availability}

The raw eMERLIN data (as described in Section~\ref{sec:eMerlinData}) is available via project code CY11205 at the following link: \url{https://www.e-merlin.ac.uk/distribute/CY11/CY11205/CY11205.html}. All the resulting maps produced for this work are available the following link: \url{https://doi.org/10.25405/data.ncl.c.7523193}. The displayed VLA maps are from \cite{Jarvis2021} and can be downloaded via \url{https://doi.org/10.25405/data.ncl.13416164} (L-band maps) and \url{https://doi.org/10.25405/data.ncl.13702021} (C-band maps).



\bibliographystyle{mnras}
\bibliography{QFS} 




\appendix




\bsp	
\label{lastpage}
\end{document}